\documentclass[twocolumn,showpacs,aps,prd]{revtex4-1}

\usepackage{atlasphysics}
\usepackage{subfigure}
\usepackage{amsmath}
\usepackage{multirow}

\usepackage{graphicx}

\newcommand\Kshort{\ensuremath{K^{0}_{\mathrm{\it S}}}\xspace}
\newcommand\LambdaZ{\ensuremath{\Lambda}\xspace}
\newcommand\LambdaZbar{\ensuremath{\overline{\Lambda}}\xspace}

\usepackage[breaklinks,colorlinks,linkcolor=blue,citecolor=blue,filecolor=blue,urlcolor=blue]{hyperref}

\begin{document}

\begin{minipage}[t]{\textwidth}
  \begin{flushright}
    CERN-PH-EP-2011-168\\
    Submitted to Phys. Rev. D
    \end{flushright}
\end{minipage}

\vspace{0.5cm}

\title{\Kshort and \LambdaZ production in $pp$ interactions at $\bf{\sqrt{s} = 0.9}$ and $\bf{7}$ \tev measured with the ATLAS detector at the LHC}
\author{G. Aad \textit{et al.}}\thanks{Full author list given at the end of the article.}
\collaboration{The ATLAS Collaboration}

\date{\today}

\begin{abstract}
The production of \Kshort and \LambdaZ hadrons is studied in $pp$
collision data at $\sqrt{s} = 0.9$ and $7$~\tev collected with the
ATLAS detector at the LHC using a minimum-bias trigger.  The observed
distributions of transverse momentum, rapidity, and multiplicity are
corrected to hadron level in a model-independent way within well
defined phase-space regions.  The distribution of the production ratio
of \LambdaZbar to \LambdaZ baryons is also measured.  The results are
compared with various Monte Carlo simulation models.  Although most of
these models agree with data to within $15\%$ in the \Kshort
distributions, substantial disagreements are found in the \LambdaZ
distributions of transverse momentum.
\end{abstract}


\maketitle

\newpage

\section{\label{secIntroduction}Introduction}

Yields and production spectra of hadrons containing strange quarks
have been measured previously at the Large Hadron Collider (LHC) and
the Tevatron at various center-of-mass
energies~\cite{ALICE_Strangeness_Paper,CMS_StrangenessPaper,CDF_StrangenessPaper}.
Measurements of particle production provide insight into the behavior
of QCD interactions at low momentum transfer, typically described by
models with empirical parameters tuned from experimental data.
Accurate modeling of such interactions is also essential for
constraining the effects of the underlying event in the high-\pt
collisions studied at the LHC.  As the strange quark is heavier than
the up and down quarks, the production of strange hadrons is
suppressed relative to hadrons containing only up and down quarks.
However, since the mass of the strange quark is comparable in value to
the $\Lambda_{\rm QCD}$ scale constant, it is not sufficiently heavy
for perturbative techniques to be used in modeling the production of
strange hadrons and experimental input is required to tune it in Monte
Carlo (MC) simulation.  Moreover, the ratio of the production of
strange antibaryons to strange baryons is related to the transfer of
baryon number from the colliding protons to the mid-rapidity region
and can be used to constrain ``diquark''~\cite{Diquarks} and
``string-junction''~\cite{StringJunction} models in MC generators.
Since the initial state in $pp$ collisions has a net baryon number of
two, these models can be tested even at zero rapidity at the LHC.

In this paper, the production of \Kshort and \LambdaZ hadrons is
studied using the first $190$~\inmb collected by the ATLAS experiment
at $\sqrt{s}=7$~\tev and $7$~\inmb at $900$~\gev.  In addition, the
measurement of the ratio between \LambdaZbar and \LambdaZ baryon
production is presented.  Data were collected with a minimum-bias
trigger with the same selection as in the inclusive minimum-bias
measurement of charged particles~\cite{ATLASMinbiasPaper}.  Strange
hadrons are reconstructed in the $\Kshort\to\pi^+\pi^-$, $\LambdaZ\to
p \pi^-$, and $\LambdaZbar\to\bar{p}\pi^+$ decay modes by identifying
two tracks originating from a displaced vertex, exploiting the long
lifetimes of strange hadrons (c$\tau\approx2.7$~cm for \Kshort hadrons
and c$\tau\approx7.9$~cm for \LambdaZ hadrons).  The measured
distributions are

\begin{equation}
\frac{1}{N} \frac{{\rm d}N}{{\rm d}\pt}\mbox{,     }\frac{1}{N} \frac{{\rm d}N}{{\rm d}y}\mbox{,     }\frac{1}{N_{\rm ev}} \frac{{\rm d}N_{\rm ev}}{{\rm d}N}{\rm , }
\end{equation}
where $N$ is the number of \Kshort or \LambdaZ hadrons, \pt is the
transverse momentum, $y$ is the rapidity~\cite{ATLASCoordinateSystem},
and $N_{\rm ev}$ is the number of events with two charged particles
satisfying $\pt>100~\mev$ and $|\eta|<2.5$.  The \LambdaZ
distributions do not include \LambdaZbar baryons, while the ratio of
$\LambdaZbar$ to $\LambdaZ$ is presented versus \pt and $y$ as a
separate measurement.  The kinematic spectra of strange hadrons are
extracted from the reconstructed distributions by correcting for
detector effects modeled with MC simulation samples that are validated
with data.  The observed distributions are corrected to the
$|\eta|<2.5$ and $\pt>100$~\mev phase-space region where tracks can be
reconstructed (imposed on the charged decay products) with minimum and
maximum flight-length requirements imposed on the \Kshort and \LambdaZ
hadrons to avoid model-dependent extrapolations outside of the
detector acceptance.  A similar approach was used in the ATLAS
measurement of charged-hadron production~\cite{ATLASMinbiasPaper}.

\section{\label{secDetector}The ATLAS detector}

The ATLAS detector~\cite{ATLASDetector} at the LHC~\cite{LHCPaper}
covers almost the whole solid angle around the collision point with
layers of tracking detectors, calorimeters and muon chambers. It has
been designed to study a wide range of physics topics at LHC
energies. For the measurements presented in this paper, the tracking
devices and the trigger system are used.

The ATLAS Inner Detector (ID) has full coverage in $\phi$ and covers
the pseudorapidity range $|\eta| < 2.5$.  It consists of a silicon
pixel detector (Pixel), a silicon microstrip detector (SCT) and a
transition radiation tracker (TRT). The sensitive elements of these
detectors cover a radial distance from the interaction point of 51-150
mm, 299-560 mm, and 563-1066 mm, respectively, and are immersed in a 2
T axial magnetic field.  The ID barrel (end-cap) region consists of 3
($2 \times 3$) Pixel layers, 4 ($2 \times 9$) double-layers of
single-sided silicon microstrips with a 40 mrad stereo angle, and 73
($2 \times 160$) layers of TRT straws. Typical position resolutions
are 10, 17 and 130 $\mu$m for the $R-\phi$ coordinate and, in the case
of the Pixel and SCT, 115 and 580 $\mu$m for the second measured
coordinate. A track from a charged particle traversing the barrel
detector would typically have 11 silicon hits (3 pixel clusters and 8
strip clusters) and more than 30 straw hits.

The ATLAS detector has a three-level trigger system; data for this
measurement were collected with Level 1 signals from the Beam Pickup
Timing devices (BPTX) and the Minimum Bias Trigger Scintillators
(MBTS). The BPTX stations consist of electrostatic button pick-up
detectors attached to the beam pipe at $\pm 175$~m from the center of
the detector. The coincidence of the BPTX signal between the two sides
of the detector is used to determine when beam bunches are colliding
in the center of the detector. The MBTS are mounted at each end of the
detector in front of the liquid-argon end-cap calorimeter cryostats at
$z = \pm 3.56$~m. They are segmented into eight sectors in azimuth and
two rings in pseudorapidity ($2.09 < |\eta| < 2.82$ and $2.82 < |\eta|
< 3.84$). Data were collected for this analysis using a trigger
requiring a BPTX coincidence and MBTS trigger signals. The MBTS
trigger used for this paper is configured to require at least one hit
above threshold from either side of the detector, referred to as a
single-arm trigger.

\section{\label{secDetectorData}Data samples and event selection}

The data used in this analysis consist of about $16$ million events
recorded by ATLAS in March and April 2010, corresponding to about
$190$~\inmb of proton--proton collisions provided by the LHC at the
center-of-mass energy of $7$~\tev, as well as $1$ million events
corresponding to about $7$~\inmb at $\sqrt{s} = 900$~\gev recorded in
December 2009. Data events are required to pass the same data-quality
and event requirements as those used in Ref.~\cite{ATLASMinbiasPaper}.
These include a primary vertex reconstructed from two or more tracks
with $\pt>100$~\mev and transverse distance of closest approach to the
beam-spot position of at most $4$~mm.  Events containing more than one
primary vertex are rejected.  After the selection, the fraction of
events with more than one interaction in the same bunch crossing in
these early LHC data is estimated to be at the $0.1\%$ level and is
neglected.

A sample of $20$ million non-diffractive minimum-bias MC events
generated with \textsc{pythia} using the early ATLAS MC09
tune~\cite{PythiaManual,Pythia} and \textsc{geant4}~\cite{GEANT}
simulation is passed through the same reconstruction as the data
sample.  The distribution of the longitudinal position of the primary
vertex in the simulated sample is re-weighted to make it consistent
with data.  Samples of single-diffractive and double-diffractive
events generated with the same tune are combined with the
non-diffractive sample according to their relative total cross
sections in the same manner as in Ref.~\cite{ATLASMinbiasPaper}.  The
distributions of the longitudinal position of the primary vertex are
found to be nearly identical in the simulated minimum-bias and
diffractive samples.  For some systematic studies, a fully simulated
sample of events produced with the \textsc{phojet}
generator~\cite{Phojet} is used.  To compare the data at particle
level with different phenomenological models describing minimum-bias
events, the following samples are also used:

\begin{itemize}
\item
\textsc{pythia6} using the AMBT2B-CTEQ6L1 tune~\cite{PythiaAMBT2B,CTEQ6L1};
\item
\textsc{pythia6} using the Perugia2011 tune~\cite{Perugia2011} (CTEQ5L
parton distribution functions (PDFs)~\cite{CTEQ5L});
\item
\textsc{pythia6} using the Z1 tune~\cite{PythiaZ1} (CTEQ5L PDFs);
\item
\textsc{pythia8} using the 4C tune\cite{Pythia8,Pythia8_2} (CTEQ6L1 PDFs);
\item
\textsc{herwig++} 2.5.1~\cite{Herwigpp,Herwigpp_2}, using the UE7-2
 underlying-event tune at $7$~\tev and the MU900-2 minimum-bias tune
 at $900$~\gev~\cite{Herwigpp_tunes} (both with MRST2007LO*
 PDFs~\cite{MRST2007LO}).
\end{itemize}

\section{\label{secSelection}V$^0$ reconstruction and selection}

\begin{figure}[!tb]
  \includegraphics[width=0.5\textwidth]{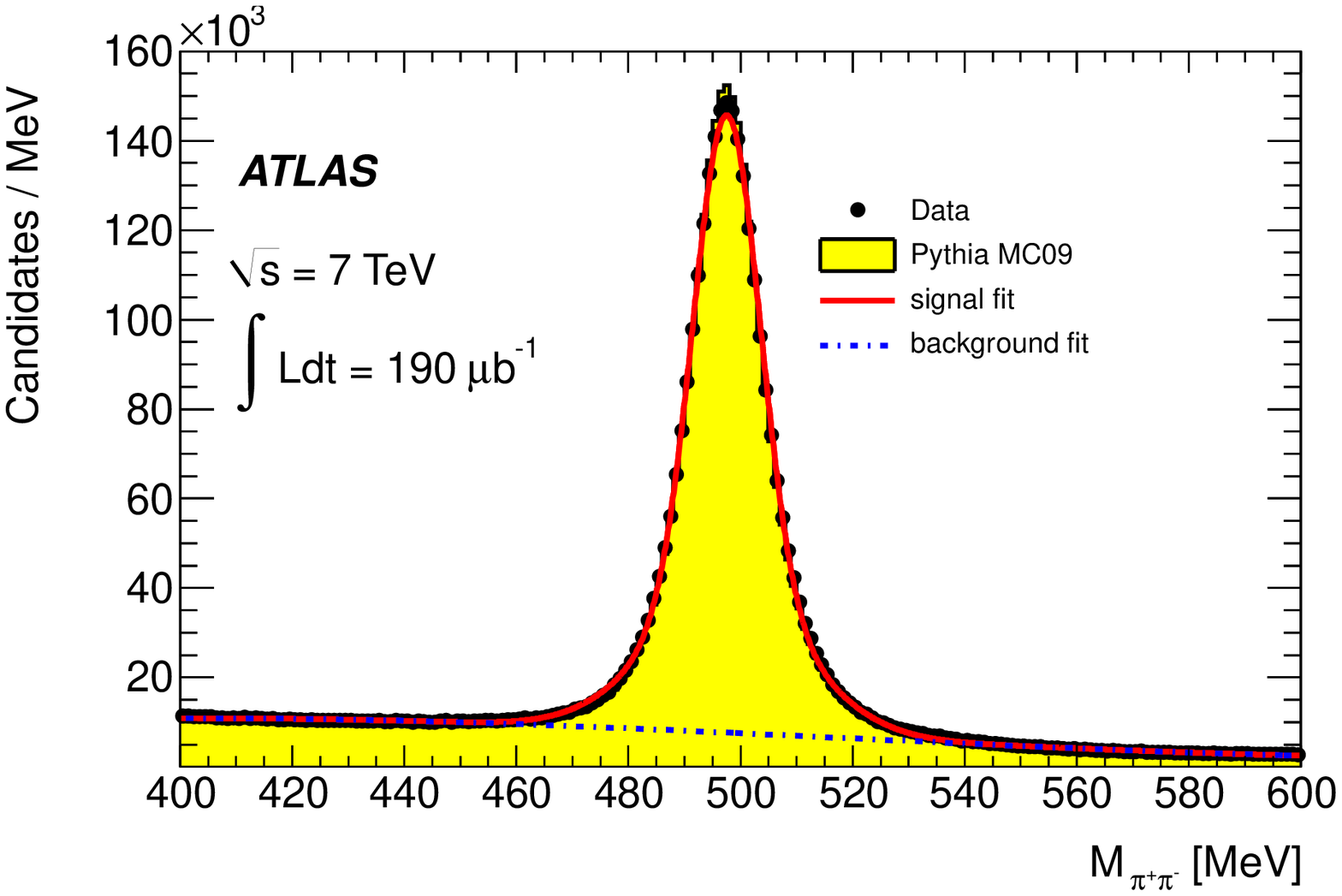}
  \includegraphics[width=0.5\textwidth]{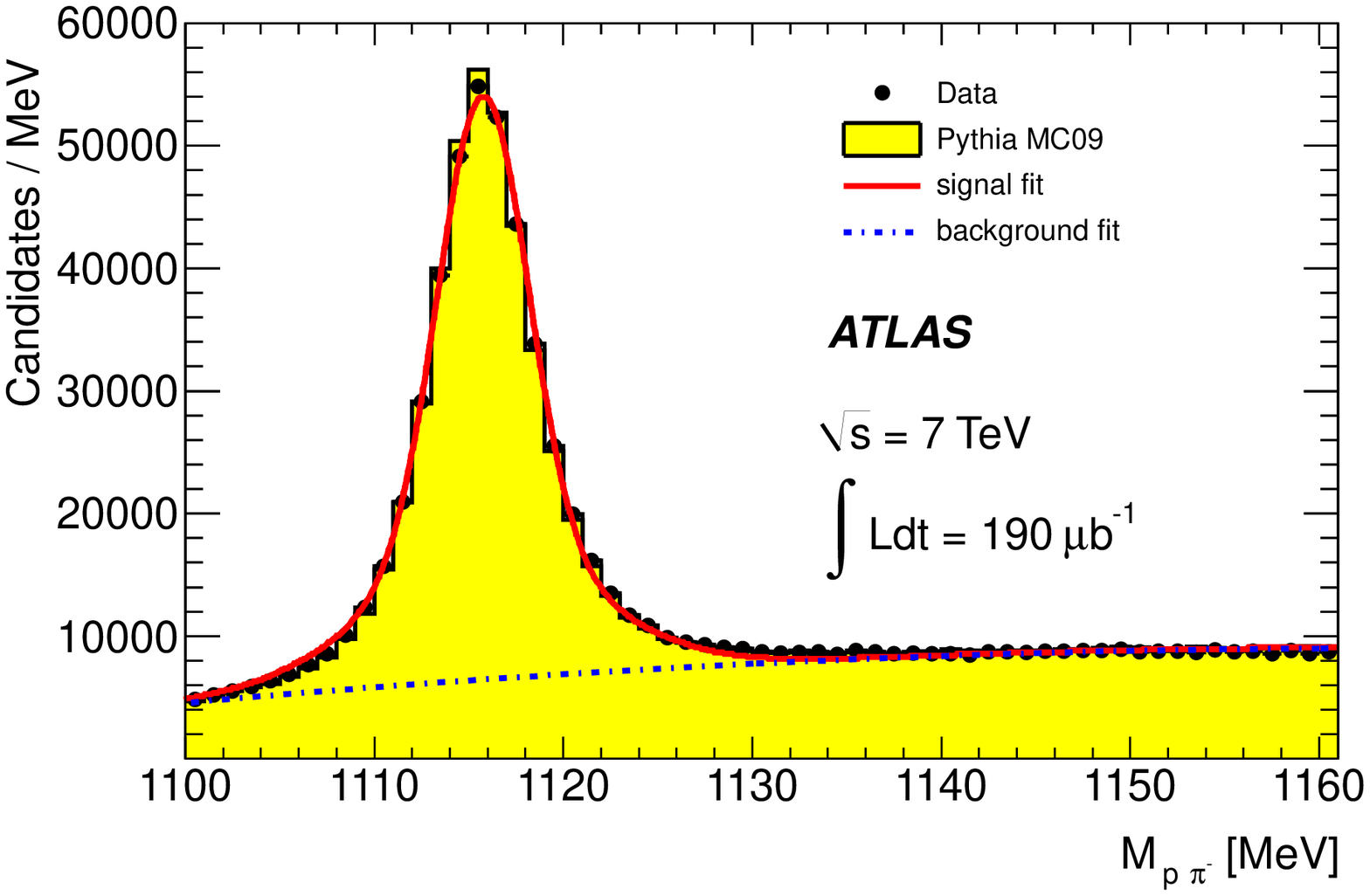}
 \caption{Comparison of measured and predicted \Kshort (top) and
 \LambdaZ (bottom) invariant-mass distributions in the $7$~\tev
 samples.  The points are data, while the histograms show the MC
 sample with signal and background components separately normalized to
 the data. The solid line is the line-shape function fitted to data,
 while the dot-dashed line shows the component of the fitted function
 describing the combinatoric background (see
 Sec.~\ref{subsecKshortBackground}).\label{fig:MassDistributions}}
\end{figure}

Tracks with $\pt>50$~\mev are reconstructed within the $|\eta|<2.5$
acceptance of the ID as described in detail in
Refs.~\cite{ATLASMinbiasPaper,PatternRecognition,TrackingAlgorithms}.
To form \Kshort candidates, oppositely charged track pairs with
$\pt>100$~\mev and at least two silicon hits are fit to a common
vertex, assuming the pion mass for both tracks.  The \Kshort
candidates are required to satisfy the following criteria:
\begin{itemize}
\item The $\chi^2$ of the two-track vertex fit is required to be less
than 15 (with 1 degree of freedom).
\item The transverse flight distance, defined by the transverse
distance between the secondary vertex (\Kshort decay point) and the
reconstructed primary vertex, is required to be between $4$~mm and
$450$~mm.
\item The cosine of the pointing angle in the transverse plane
($\cos{\theta_{\rm K}}$) between the \Kshort momentum vector and the
\Kshort flight direction, defined as the line connecting the
reconstructed primary vertex to the decay vertex, is required to be
greater than 0.999 (equivalent to an angle of $2.56^{\circ}$).
\end{itemize}
For \LambdaZ and \LambdaZbar decays, the track with the higher \pT is
assigned the proton mass and the other track is assigned the pion
mass.  In the simulated sample this identification is correct for
$99.8\%$ of the candidates.  The \LambdaZ and \LambdaZbar candidates
are required to satisfy the following criteria:
\begin{itemize}
\item The $\chi^2$ of the two-track vertex fit is required to be less
than 15 (with 1 degree of freedom).
\item The transverse flight distance is required to be between $17$~mm
and $450$~mm.
\item The cosine of the pointing angle is required to be greater than
  0.9998 (equivalent to an angle of $1.15^{\circ}$).
\item The \pt of the \LambdaZ candidate is required to be greater than
$500$~\mev
\end{itemize}
These requirements reduce the combinatorial background.  The smaller
signal-to-background ratio in the \LambdaZ sample with respect to the
\Kshort sample requires a tighter pointing requirement, while the
larger value of the flight-distance selection exploits the longer
lifetime of the \LambdaZ baryon.  The minimum \pt cut removes poorly
reconstructed candidates.  The distributions of the invariant mass of
the \Kshort and \LambdaZ candidates in the data and MC samples are
shown in Fig.~\ref{fig:MassDistributions}.

\begin{figure}[!tb]
  \includegraphics[width=0.5\textwidth]{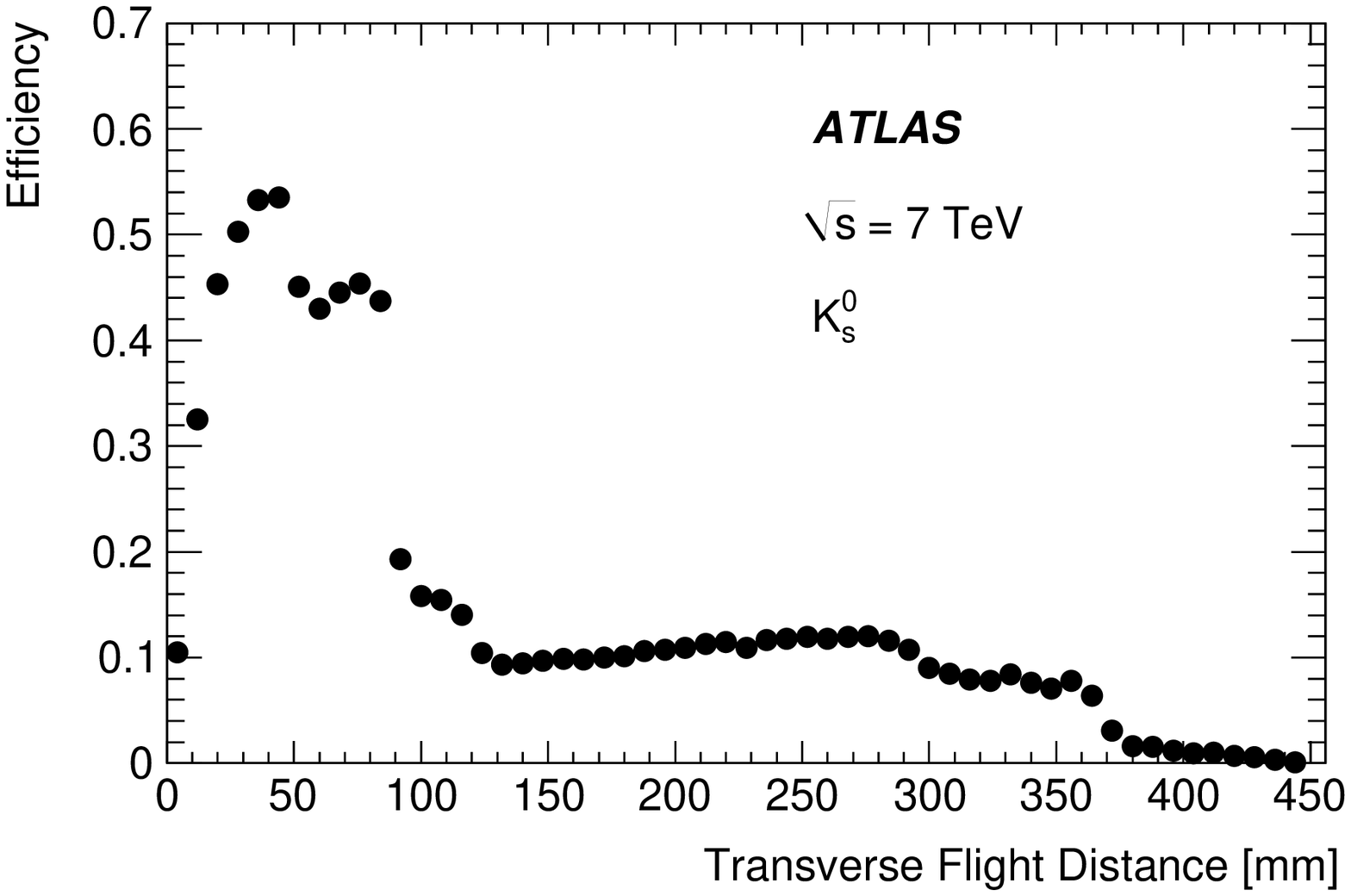}
  \includegraphics[width=0.5\textwidth]{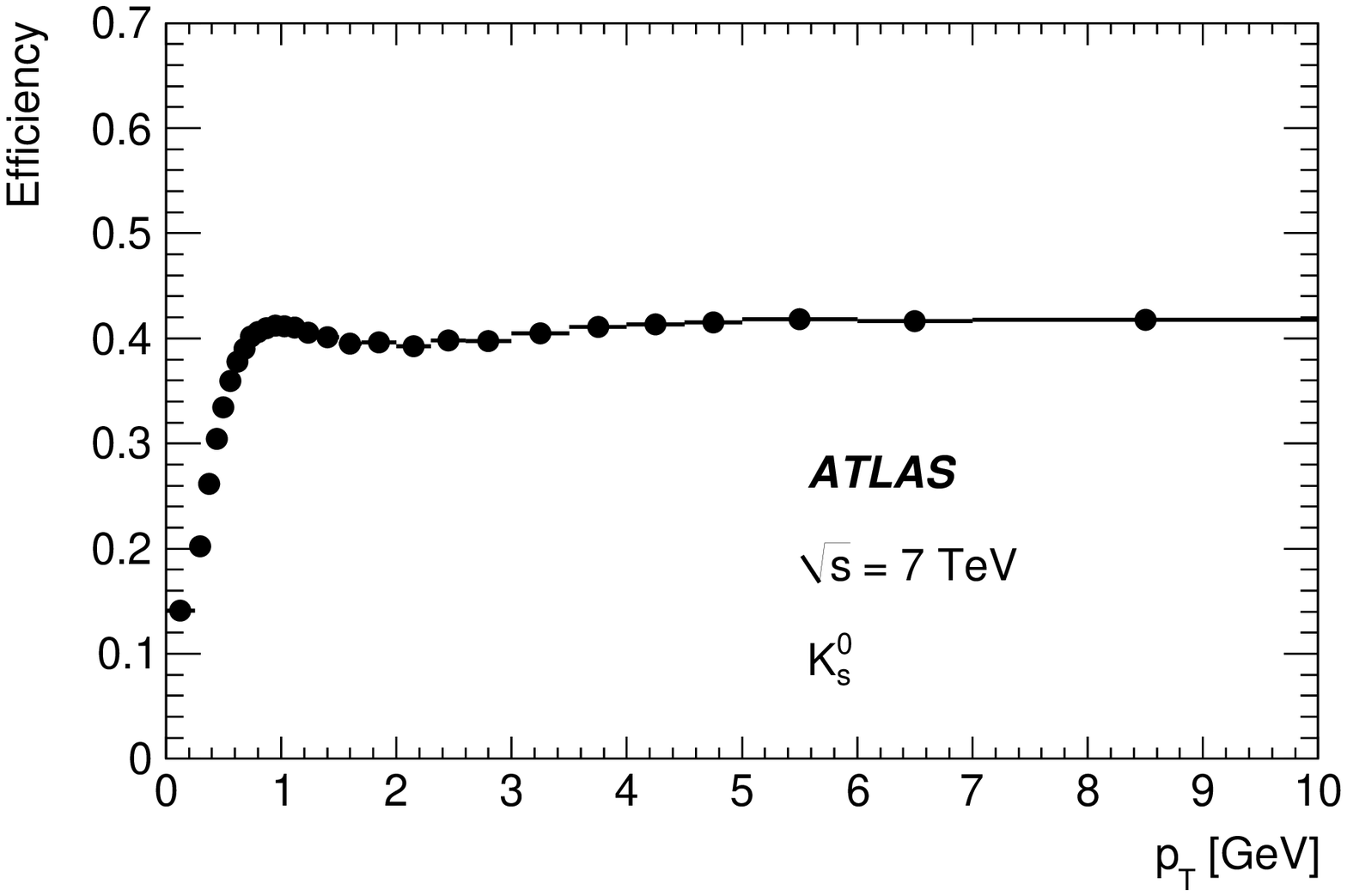}
  \includegraphics[width=0.5\textwidth]{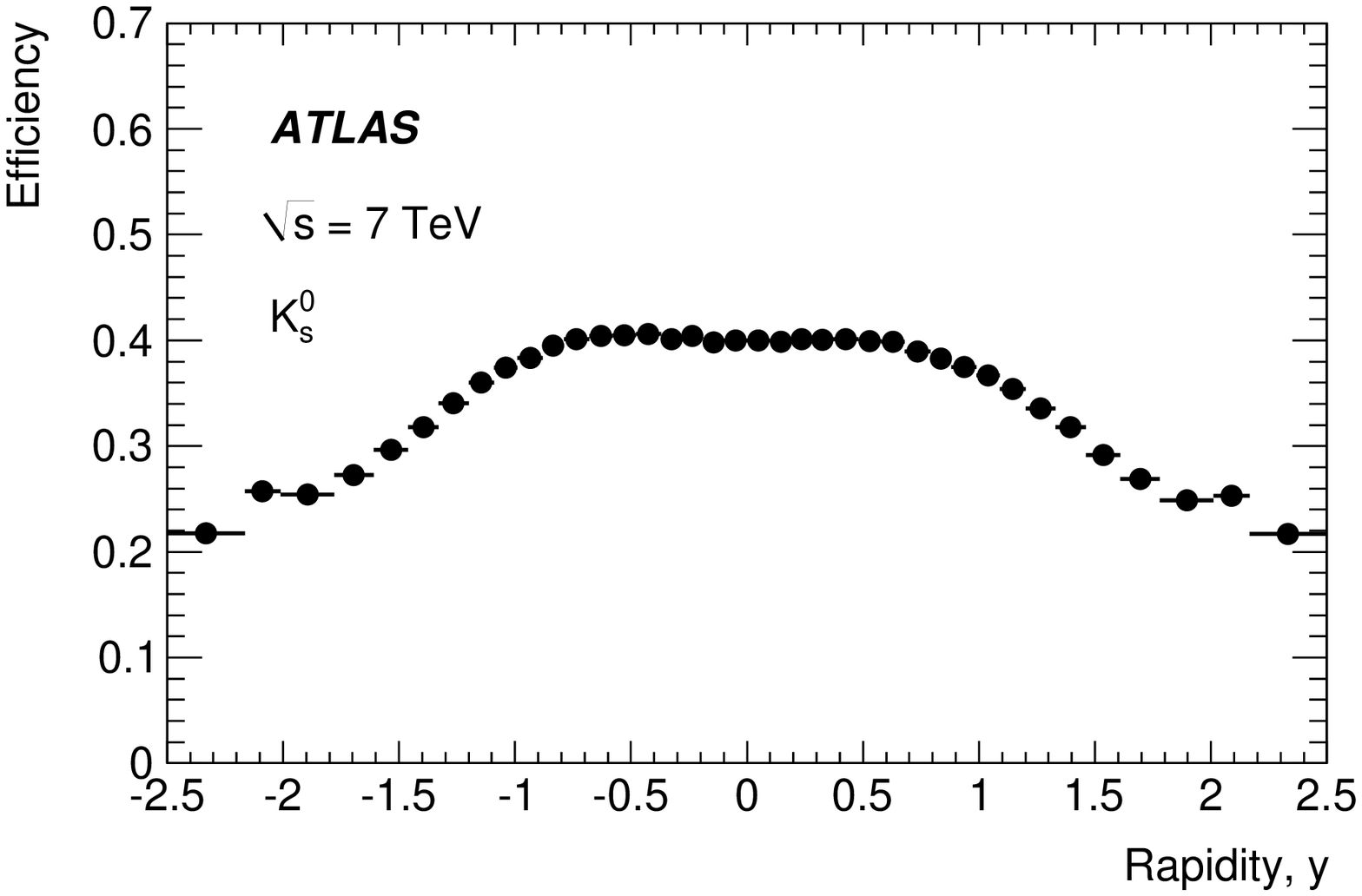}
  \vspace{-0.5cm}
  \caption{The reconstruction efficiency of \Kshort candidates in the
   $7$~\tev MC sample after all selection criteria versus the
   transverse flight distance (top), \pt (center), and rapidity
   (bottom).\label{fig:EfficiencyRadiusPt}}
\end{figure}

Figures~\ref{fig:EfficiencyRadiusPt} and \ref{fig:RatioEfficiency}
show the reconstruction efficiency of \Kshort, \LambdaZ, and
\LambdaZbar candidates versus the radial position of the decay vertex,
\pt, and rapidity.  The efficiency is determined from simulation by
comparing the number of generated \Kshort hadrons with the number of
reconstructed candidates after all selection criteria are applied.
The efficiency turn-on curve versus \pt is mainly an effect of
tracking efficiency, while the radial plot clearly shows the drops in
efficiency when crossing detector layers, reflecting the lower
efficiency of reconstructing and selecting tracks that have fewer hits
in the silicon detector.  (The effect is most pronounced at the Pixel
layers, located roughly at radii of $50$, $80$, and $120$~mm.)

\section{\label{secEffCorr}Efficiency and correction procedure}

The measured \Kshort and \LambdaZ production quantities are
distributions versus rapidity and transverse momentum as well as the
number of \Kshort or \LambdaZ candidates per event (the
``multiplicity''). To remove the background from the \pt and rapidity
distributions, the reconstructed invariant-mass distribution is fitted
for signal and background separately in every bin of \pt and rapidity.
The background-subtracted distributions are then corrected through an
unfolding algorithm for detector resolution of the \pt and rapidity
measurements as well as for the reconstruction efficiency.  In the
measurement of the production ratio of \LambdaZbar to \LambdaZ
baryons, a separate correction procedure is employed accounting for
the difference in the detector response to positively and negatively
charged baryons.

\subsection{\label{subsecKshortCorr}Corrections to \Kshort and \LambdaZ distributions}

The corrections are evaluated separately for the $7$~\tev and
$900$~\gev samples and are described sequentially below.  The final
distributions are normalized to unity by dividing by the total number
of measured hadrons.

\subsubsection{\label{subsecKshortBackground}Background correction}

The number of signal candidates in a given bin of the rapidity and
transverse-momentum distributions is determined by fitting the
invariant-mass spectrum of the \Kshort or \LambdaZ candidates in that
bin.  The value and statistical uncertainty on the bin are then
determined from the fitted signal yield and its uncertainty.  For the
\Kshort candidates the functional form that is found to describe well
the shape in data combines the sum of two Gaussians for the signal
peak and a third-order polynomial for the combinatorial
background. The means of the two Gaussian components are constrained
to be the same, while the widths and relative fractions are determined
from the fit.  For the \LambdaZ candidates a second-order polynomial
is used for the background and the following modified Gaussian shape
is used for the signal:
\begin{equation}
C \cdot \exp\left[ -0.5\cdot x^{\left(1+\frac{1}{1+0.5\cdot
x}\right)}\right] {\rm , } \: x=\left| \frac{m-\mu}{\sigma} \right|
{\rm ,}
\end{equation}
where $m$ is the invariant mass and the fitted parameters are the
normalization parameter $C$, the mean $\mu$, and the width $\sigma$.
This shape is found to model the invariant mass better than the sum of
two Gaussians.

The results of the fits to the entire $7$~\tev data and MC samples are
summarized in Table~\ref{FitResults}. The means of the mass peaks
obtained from the fits in data are in reasonable agreement with
simulation and with the world average~\cite{PDGRef}.  The agreement
demonstrates the accuracy of the track momentum scale and of the
modeling of the Inner Detector's $2$\,T solenoid magnetic field, which
has been mapped to a precision of about $0.4$\,mT~\cite{ATLASMagnet}.
Although the deviation of data from the simulated and world-average
values is statistically significant since the uncertainties do not
include systematic effects, it is no larger than about $100$~\kev and
does not affect the results presented in this article, as the mean
mass position is not directly used in the measurement.

The contamination from secondary \Kshort and \LambdaZ production from
long-lived baryon decays or nuclear interactions in the detector
material is at the negligible level of $0.1\%$ for \Kshort decays in
simulation and at the $10\%$ level in the \LambdaZ case, where it is
subtracted from the measured data distributions.  The modeling of
secondary \LambdaZ baryons is evaluated by varying the pointing-angle
selection and comparing its efficiency between MC and data.  The
measured deviations at the level of $2\%$ in the efficiency are
assessed as a systematic uncertainty.  The effect of \LambdaZ
contamination in the \Kshort signal and vice versa is similarly
studied and the contamination of less than $1\%$ is included in the
evaluation of systematic uncertainties.

\begin{table}
\begin{center}
\begin{tabular}{|l|c|c|}
  \hline
 & fit mean [MeV] & world average [MeV] \\
\hline
\hline
\Kshort Data& 497.536 $\pm$ 0.006 & 497.614 $\pm$ 0.024 \\
  \cline{1-2}
\Kshort MC& 497.495 $\pm$ 0.006 & \\
  \hline
  \hline
\LambdaZ Data& 1115.75 $\pm$ 0.01 & 1115.683 $\pm$ 0.006    \\
  \cline{1-2}
\LambdaZ MC & 1115.72 $\pm$ 0.01 &     \\
  \cline{1-2}
\LambdaZbar Data& 1115.81 $\pm$ 0.01 &    \\
  \cline{1-2}
\LambdaZbar MC& 1115.76 $\pm$ 0.01 &  \\
\hline
\end{tabular}
\end{center}
\caption{The position of the mass peak in the fit to the $7$~\tev data
and simulation samples.  The fit uncertainties on the mean are
statistical only.}
\label{FitResults}
\end{table}

\subsubsection{\label{subsecKshortResolution}Resolution correction}

The \textsc{pythia} MC09 simulation sample is used to fill a
two-dimensional migration matrix, where one dimension is binned in the
generated value of the variable of interest (\pt, rapidity, or
multiplicity) and the other is binned in the reconstructed value of
the same variable.  This matrix thus models the effect of the
experimental resolution on the true value of \pt or rapidity for
reconstructed candidates, which are matched to the generated
candidates using a hit-based matching
algorithm~\cite{TrackingAlgorithms}.  This matrix is then used to
unfold the migration across bins in the background-subtracted
distributions in data.

\subsubsection{\label{subsecKshortEfficiency}Efficiency correction}

The resolution-corrected \pT and rapidity distributions from the
previous step are corrected bin by bin for the reconstruction
efficiency, $\epsilon_i$, in a given bin $i$.  The correction factor,
$1/\epsilon_i$, is derived from the \textsc{pythia} MC09 sample as the
ratio of generated to reconstructed candidates in bin $i$ of the
generated distribution.  Only the generated \Kshort and \LambdaZ
hadrons originating from the primary vertex and decaying within the
tracking acceptance are considered: the two pions (the proton and the
pion) that the \Kshort (\LambdaZ) hadron decays to are required to
have $|\eta|<2.5$ and $\pt>100\mev$, while the \Kshort or \LambdaZ
hadron itself is required to satisfy the appropriate minimum
flight-distance requirement and a maximum flight-distance requirement
of $450$~mm, which corresponds to the effective acceptance imposed by
the silicon hit-content selection on the tracks.  The reconstructed
distributions in data are thus corrected to particles produced within
the same acceptance, as extrapolating to regions not probed by the
Inner Detector would introduce a dependence on the MC generator model
in the correction procedure.  The efficiency derived from MC is binned
in \pT or rapidity and the effectiveness of the entire correction
procedure is evaluated through pseudo-experiments where the
\textsc{phojet} MC sample is unfolded using migration matrices filled
from the \textsc{pythia} MC09 sample.  (See
Section~\ref{secSystematics}.)

\subsection{\label{subsecLambdaCorr}Corrections to the \LambdaZbar/\LambdaZ production ratio}

\begin{figure}[!tb]
\begin{center}
  \includegraphics[width=0.5\textwidth]{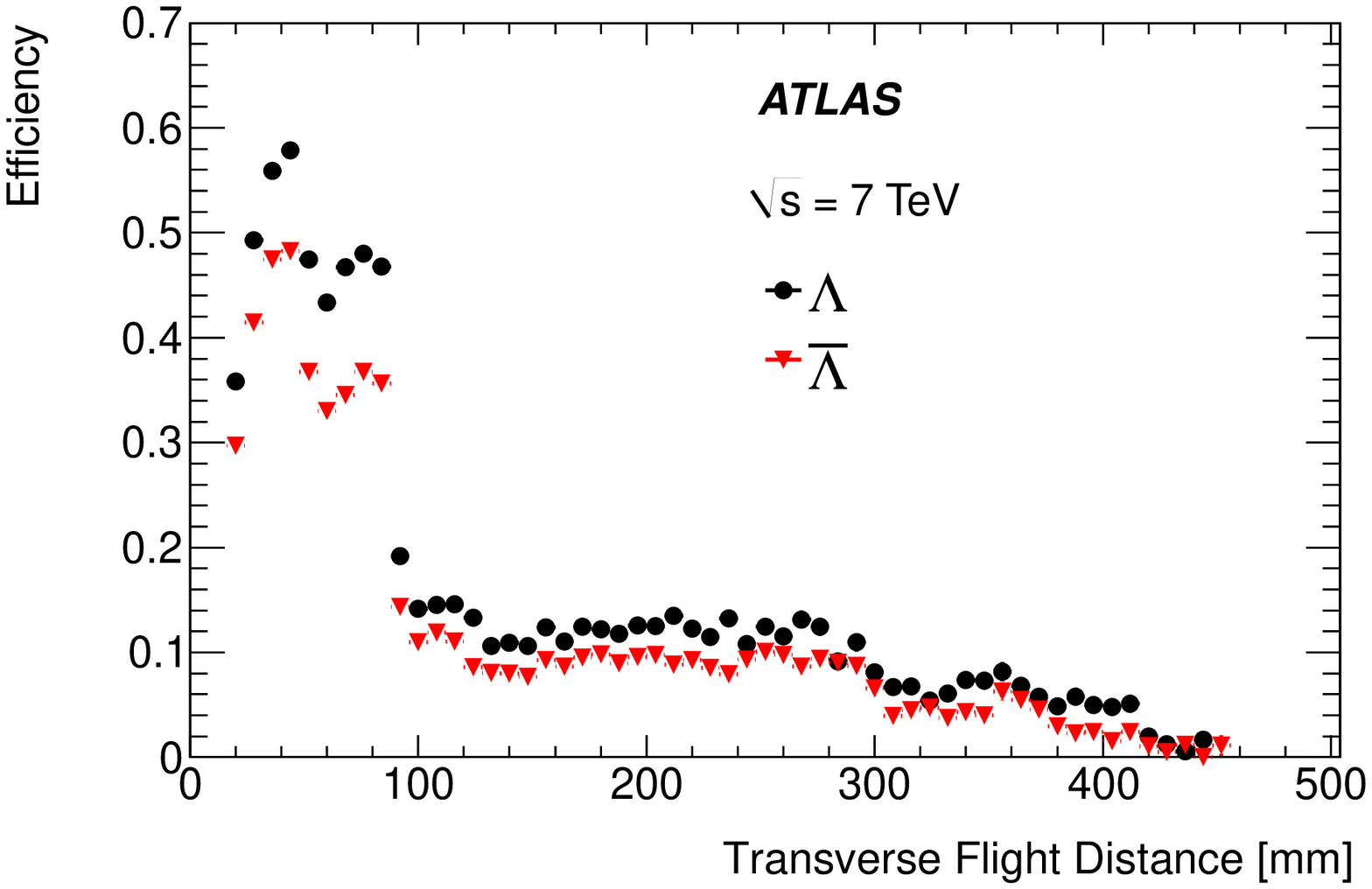}
  \includegraphics[width=0.5\textwidth]{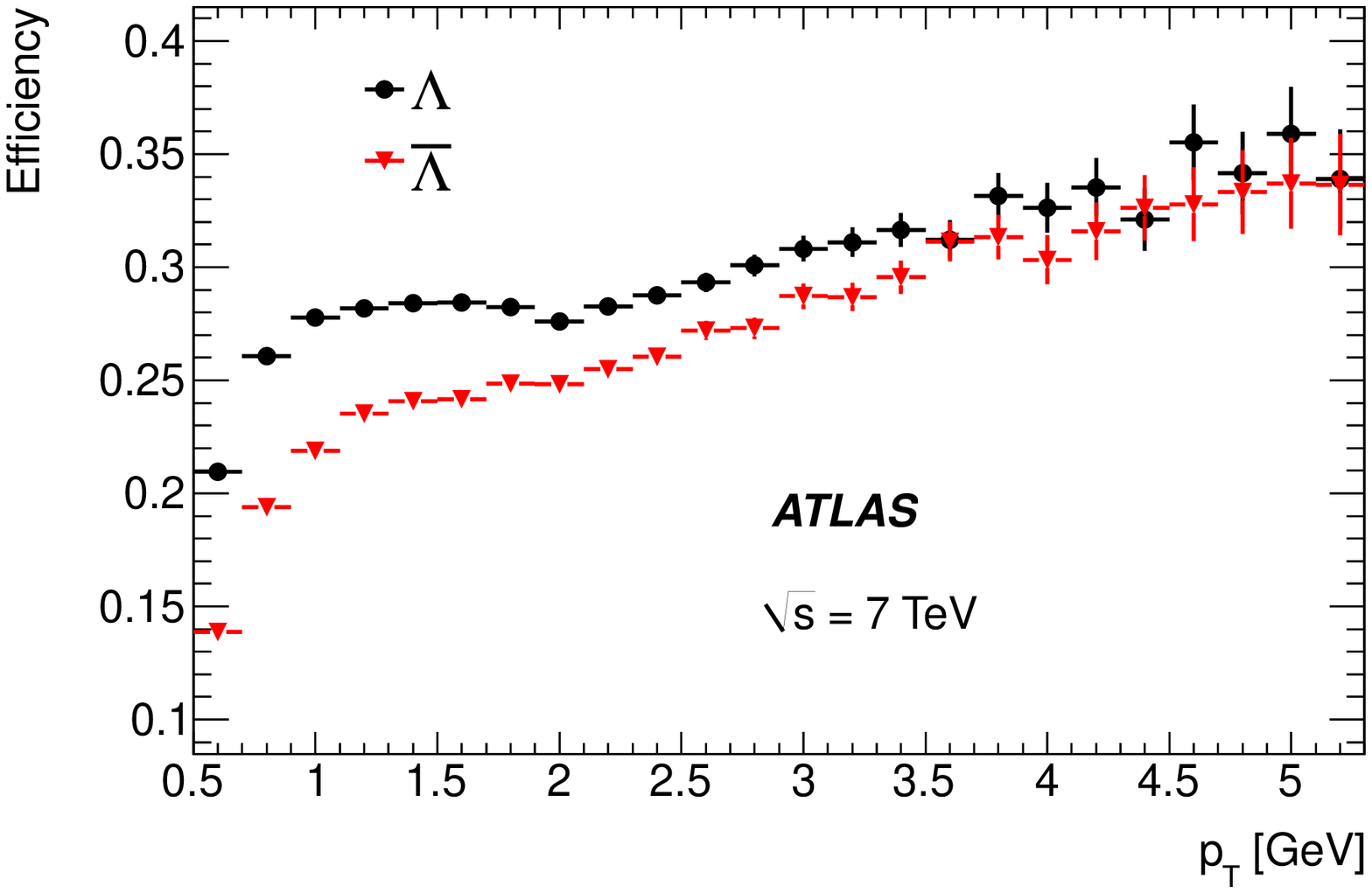}
  \includegraphics[width=0.5\textwidth]{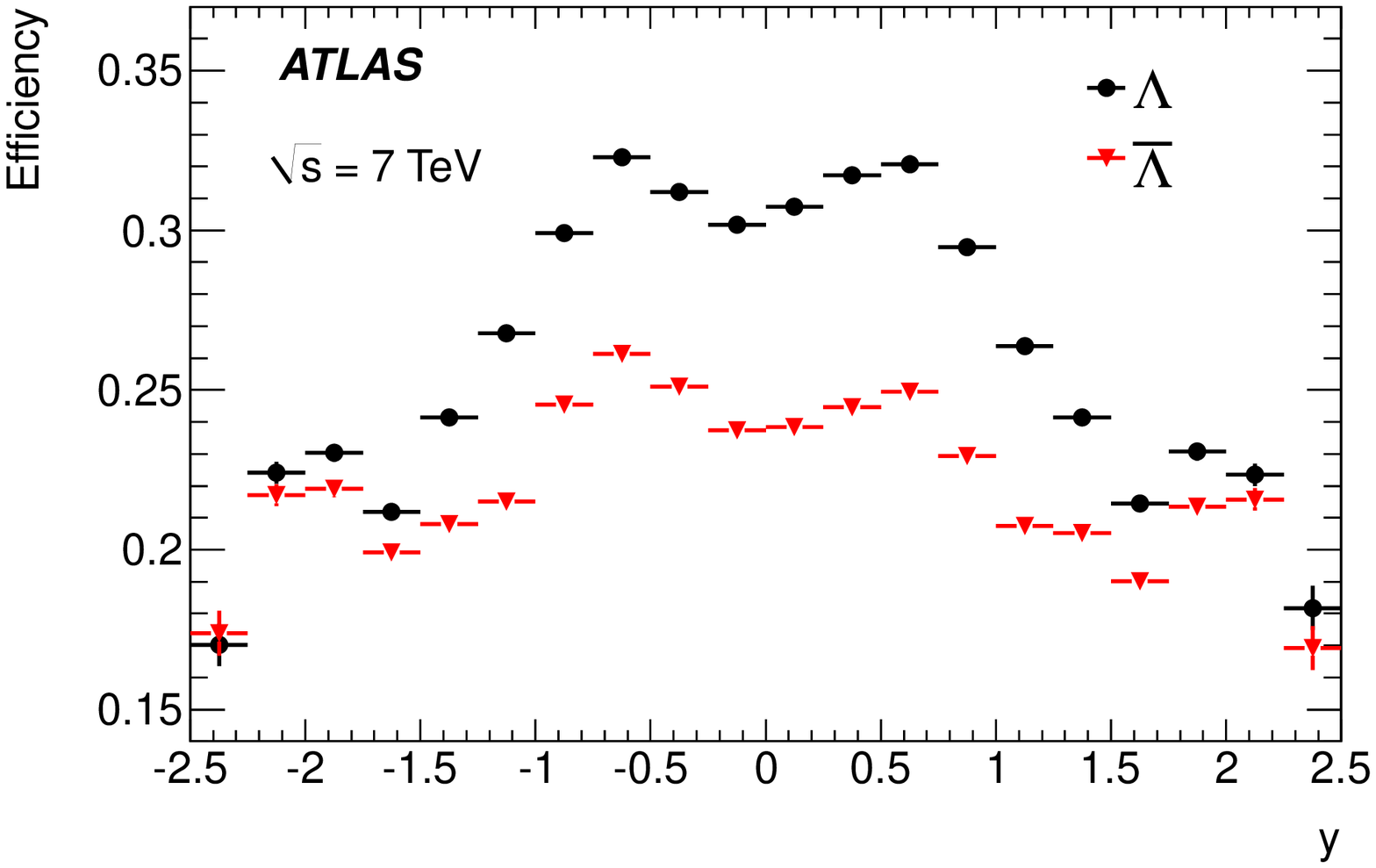}
  \vspace{-0.5cm}
  \caption{The efficiency in $7$~\tev MC for reconstructing
  \LambdaZbar and \LambdaZ candidates after all selection criteria
  versus the transverse flight distance (top), \pt (middle) and
  rapidity (bottom).  The uncertainties are statistical
  only.\label{fig:RatioEfficiency}}
\end{center}
\end{figure}

The background in the \LambdaZbar and \LambdaZ distributions is
subtracted in the same manner as the \Kshort background but with the
modified Gaussian shape for the signal component.  As most systematic
tracking effects cancel in the production ratio, the ratio is
corrected only for the difference in reconstruction efficiency between
\LambdaZ and \LambdaZbar decays.  This difference is mainly a
consequence of the difference in tracking efficiency between protons
(for \LambdaZ candidates) and antiprotons (for \LambdaZbar candidates)
caused by different interactions with detector material.  The
correction is estimated from the MC sample in bins of \pt and rapidity
by comparing the reconstruction efficiency for \LambdaZ and
\LambdaZbar decays, which is shown in Fig.~\ref{fig:RatioEfficiency}.
The ALICE experiment has reported that the nuclear-interaction cross
section of antiprotons used by \textsc{geant4} is
over-estimated~\cite{ALICE_Strangeness_Paper, ALICE_Proton_Paper},
resulting in an over-estimated efficiency difference between \LambdaZ
and \LambdaZbar reconstruction as shown in
Fig.~\ref{fig:RatioEfficiency}.  Validation and correction of the
model of detector material and the \textsc{geant} modeling of
material-interaction cross sections and the associated systematic
uncertainties are described in Section~\ref{secSystematics}.

\section{\label{secSystematics}Systematic uncertainties}

The systematic uncertainties are evaluated separately for the
measurement of the \Kshort and \LambdaZ distributions and for the
measurement of the \LambdaZbar/\LambdaZ production ratio.  For the
\Kshort and \LambdaZ distributions, systematic uncertainties are
evaluated for the reconstruction efficiency, the
background-subtraction procedure, the method of correcting for the
resolution and efficiency, and the event selection.  For the
measurement of the \LambdaZbar/\LambdaZ production ratio, the modeling
of proton and antiproton reconstruction, the effect of \LambdaZ
baryons interacting with the detector material before decaying, and
the production of secondary \LambdaZ baryons are considered.

\subsection{\label{subsecEfficiencySyst}Reconstruction efficiency}

The systematic uncertainty on the efficiency is evaluated by comparing
impact-parameter distributions between the MC and data samples.  This
uncertainty is then cross-checked by comparing decay-time
distributions with the lifetime of \Kshort mesons and comparing the
selection efficiencies between MC and data.

\begin{figure}[!t]
  \includegraphics[width=0.51\textwidth]{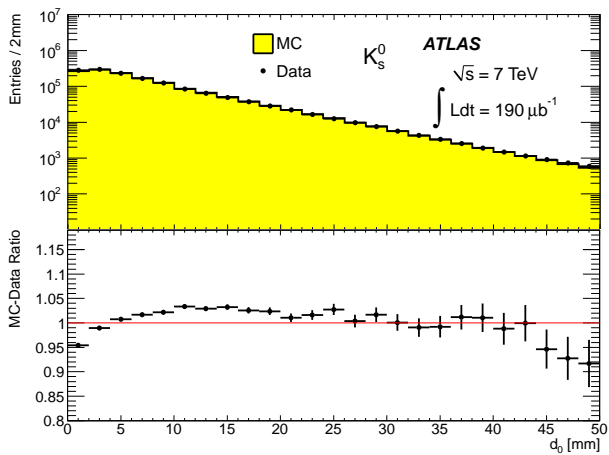}
  \vspace{-0.55cm}
  \caption{ The distribution of the reconstructed transverse impact
    parameter in $7$~\tev data and MC for pions originating in \Kshort
    decays after all selection criteria are imposed.
    \label{fig:IPdistribution}}
\end{figure}

\begin{figure}[!t]
  \includegraphics[width=0.51\textwidth]{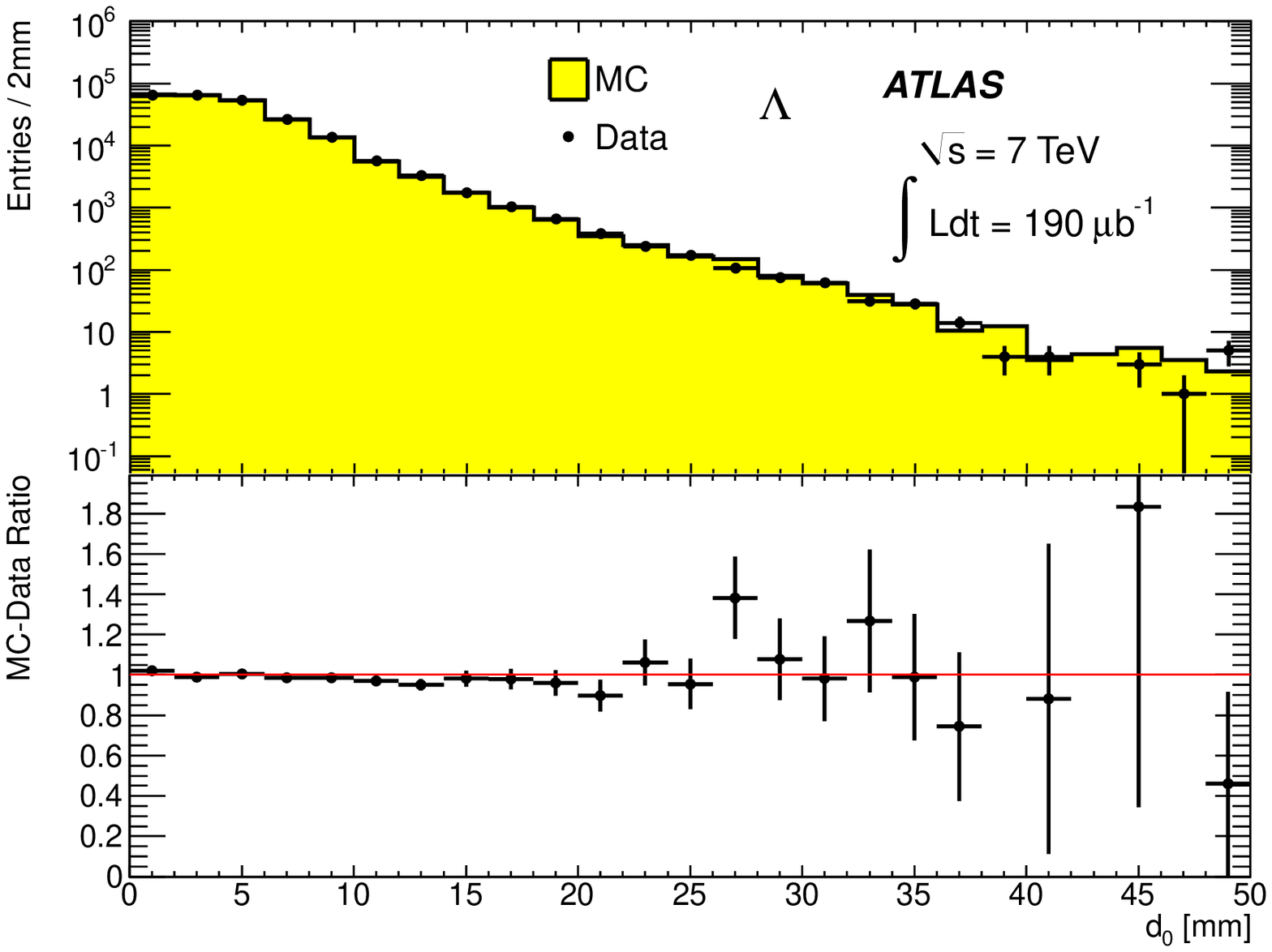}
  \includegraphics[width=0.51\textwidth]{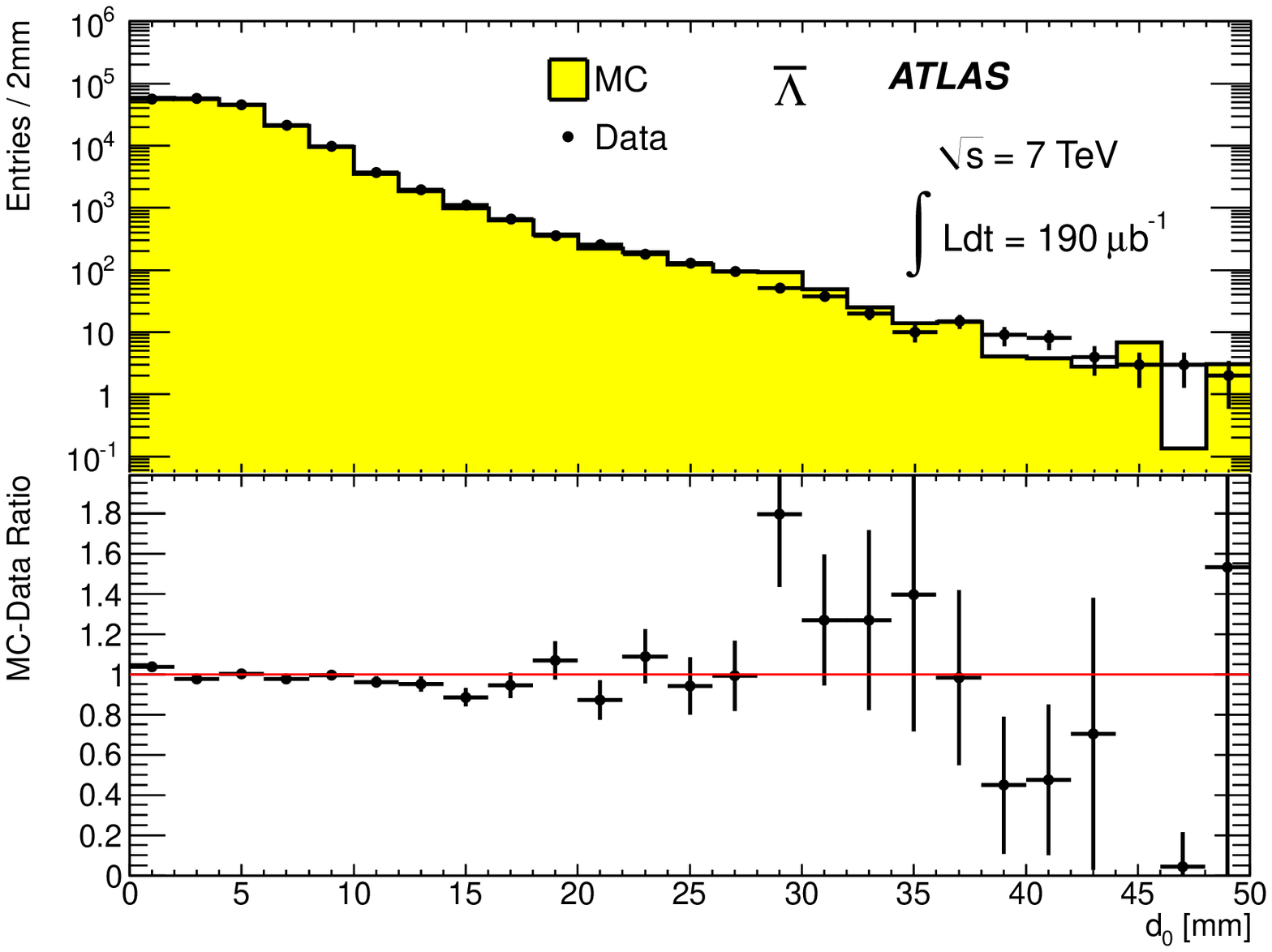}
  \vspace{-0.55cm}
  \caption{ The distribution of the reconstructed transverse impact
    parameter in $7$~\tev data and MC for protons and antiprotons
    originating in \LambdaZ (top) and \LambdaZbar decays (bottom) with
    \pt$>500\mev$ after all selection criteria are imposed.
    \label{fig:IPdistributionLambda}}
\end{figure}

\subsubsection{\label{subsubsecD0Syst}Impact-parameter distributions}

The systematic uncertainty on the tracking efficiency is evaluated
using the transverse impact parameter, $d_0$, of the tracks produced
in the \Kshort or \LambdaZ decay.  The $d_0$ measurement is sensitive
to different orientations of tracks with respect to the primary vertex
and it is correlated with the measured flight distance of the \Kshort
candidate through the vertexing of the decay point.
Figures~\ref{fig:IPdistribution} and~\ref{fig:IPdistributionLambda}
show a comparison of the reconstructed $d_0$ distributions in the data
and MC samples.

In a given two-dimensional \pt-rapidity bin, the $d_0$ distribution in
the MC sample is normalized to data.  The absolute values of the
deviations between data and MC for all $d_0$ bins are summed,
corrected for the expected value from statistical fluctuations, and
divided by the integral of the distribution.  This summed relative
difference is then assigned as the relative systematic uncertainty on
the efficiency in that \pt-rapidity bin.  The two-dimensional
\pt-rapidity uncertainty map is then projected onto each axis to
determine the one-dimensional uncertainty on the efficiency versus
either \pt or rapidity.  The uncertainty for the \Kshort efficiency is
at the $1\%$ level or less in the \pt projection except at high-\pt,
where the deviation increases to $5\%$, and at around $200$~\mev,
where it rises to $3\%$.  When evaluated versus rapidity, the typical
uncertainty is $1\%$.  The corresponding uncertainty versus rapidity
for the \LambdaZ candidates is at $2\%$, with larger uncertainties at
low \pt.  The effect of the uncertainty in the detector material on
the $d_0$ distribution in the simulation is also studied and verified
to be consistent with the results of previous studies of detector
material in minimum-bias events~\cite{ATLASMinbiasPaper}.

\subsubsection{\label{subsubsecLifetimeSyst}Decay-time distributions}

The distribution of the \Kshort proper decay time is used to
cross-check the modeling of the reconstruction efficiency in MC
simulation.  This method is sensitive to the variation of efficiency
versus flight distance and \pt, as both are correlated with the decay
time.  The background-subtracted decay-time distribution in data is
unfolded in the same manner as the \pt and rapidity distributions,
accounting for bin migration and efficiency separately according to
the MC corrections.  The unfolded distribution in data is then fitted
with an exponential shape and the lifetime compared with the
world-average value.  The fitted value of the lifetime, $89.37 \pm
0.13$~ps, is consistent with the world-average value of $89.58$~ps to
better than $0.3\%$, indicating excellent modeling of the variation of
tracking efficiency versus flight distance.

\subsubsection{\label{subsubsecCutSelectionSyst}Selection requirements}

Although the previous two methods already include systematic
uncertainties due to the flight-distance and kinematic selection
criteria, the separate systematic effect of the selection requirements
is studied as an additional cross-check on the reconstruction
efficiency; the result of this study is not included in the total
uncertainty.  The signal efficiency of each criterion is evaluated by
fitting the invariant-mass distribution before and after the selection
is imposed in the same manner as in the background subtraction, with
all other selection criteria already applied. The difference between
the data and MC samples in the value of this efficiency is taken as a
measure of how accurately the selection is modeled in the MC sample.
The deviation is evaluated in bins of \pt and rapidity, with the
finest granularity allowed by the stability and precision of the
fitting procedure.  For the silicon hit-content, flight-distance,
track-momentum, and $\chi^2$ requirements, the deviation is at the
$1\%$ level in most bins and under $2\%$ in all bins.  For the
pointing-angle requirement, the deviation is at the $2\%$ level in
most regions, but can reach higher levels in a few bins in regions of
large material and at low \pt.  These systematic effects due to the
selection requirements are consistent with the quoted systematic
uncertainties obtained from the impact-parameter study.

\subsection{\label{subsecBackgroundSyst}Background}

The systematic uncertainty on the background subtraction is evaluated
by comparing the signal yield from the fit to the invariant-mass
distribution with the number obtained by simple sideband subtraction.
The deviation for the \Kshort candidates is at the $1\%$ level in the
barrel rapidity region and rises to roughly $4\%$ in the forward
rapidity region, as can be seen in
Fig.~\ref{fig:AllSystematicsKshorts}.  The uncertainty for the
\LambdaZ candidates is roughly twice as large, as can be seen in
Fig.~\ref{fig:AllSystematicsLambdas}, reflecting the smaller
signal-to-background levels.  The $2\%$ uncertainty due to secondary
\LambdaZ production is also included in
Fig.~\ref{fig:AllSystematicsLambdas}.

\subsection{\label{subsecUnfoldingSyst}Correction procedure for resolution and efficiency}

To test the accuracy of the unfolding procedure, the reconstructed \pt
and rapidity distributions in the \textsc{phojet} MC sample are
unfolded using the corrections derived from the \textsc{pythia} MC
sample.  As the difference between the \textsc{phojet} and
\textsc{pythia} distributions is larger than the difference between
the \textsc{pythia} and data distributions, this is a conservative
test of any model dependence in the unfolding procedure. To remove the
effect of statistical fluctuations, the reconstructed distribution in
the \textsc{phojet} sample is used to generate 10000
pseudo-experiments by Poisson variation of each bin.  The
pseudo-experiments are then unfolded and the residual distribution for
each \pt or rapidity bin with respect to the particle-level
distribution in the \textsc{phojet} sample is fitted to a Gaussian
shape.  The fitted residual mean is an indication of the bias due to
the unfolding procedure in the bin, while the width is an estimate of
the statistical uncertainty on the unfolding.  The bias is at the
$3\%$ level or less in most \Kshort rapidity bins and at the $5\%$
level in the \pt bins with most of the \Kshort candidates.  For the
\LambdaZ candidates, the bias is at the $8\%$ level in most rapidity
bins and at the $5\%$ level in the \pt bins with most of the
candidates.  These biases are assigned as the systematic uncertainty
on the unfolding procedure.  The bias due to unfolding the
multiplicity distribution is evaluated in a similar manner, with the
resulting uncertainty rising with multiplicity and reaching the $20\%$
level in the three-candidate bin in the \Kshort case and $40\%$ in the
\LambdaZ case.

The statistical uncertainty on the corrected distributions in data is
evaluated from the spread in the residual distribution when unfolding
$10000$ pseudo-experiments generated from the reconstructed data
distributions.  These uncertainties include both the fluctuations in
the reconstructed distribution itself and any statistical spread from
the correction procedure.

\subsection{\label{subsecEventSyst}Event selection}

As the data sample and event selection requirements in this
measurement are identical to those used in
Ref.~\cite{ATLASMinbiasPaper}, the systematic uncertainties on the
event selection are taken directly from that analysis.  These include
uncertainties on the presence of beam backgrounds, the trigger
efficiency, the efficiency of primary vertexing, and the presence of
additional primary vertices from pile-up collisions.  The total
systematic uncertainty on the number of \Kshort and \LambdaZ hadrons
due to the event selection is $0.1\%$.

\subsection{\label{subsecTotalKshortSyst}Total uncertainty on \Kshort and \LambdaZ production}

\begin{figure}[!tb]
\begin{center}
  \includegraphics[width=0.5\textwidth]{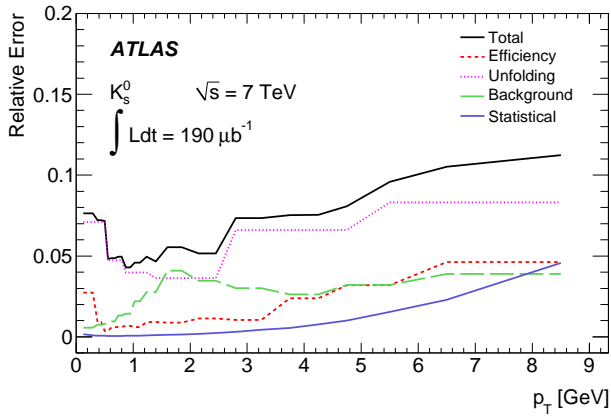}
  \includegraphics[width=0.5\textwidth]{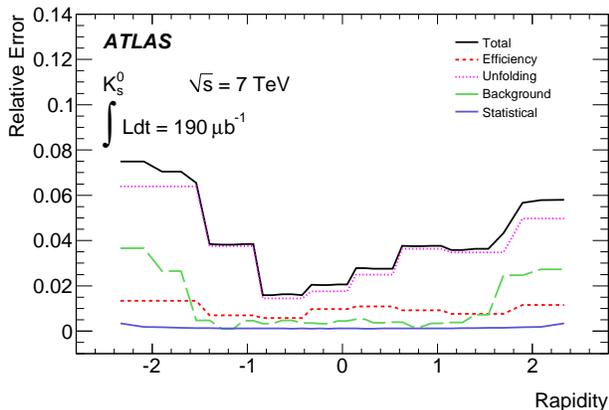}
 \vspace{-1.cm}
  \caption{ The systematic, statistical, and total uncertainties
  versus \pt (top) and rapidity (bottom) of the \Kshort candidate in
  $7$~\tev data.
    \label{fig:AllSystematicsKshorts}}
\end{center}
\end{figure}

\begin{figure}[!tb]
\begin{center}
  \includegraphics[width=0.5\textwidth]{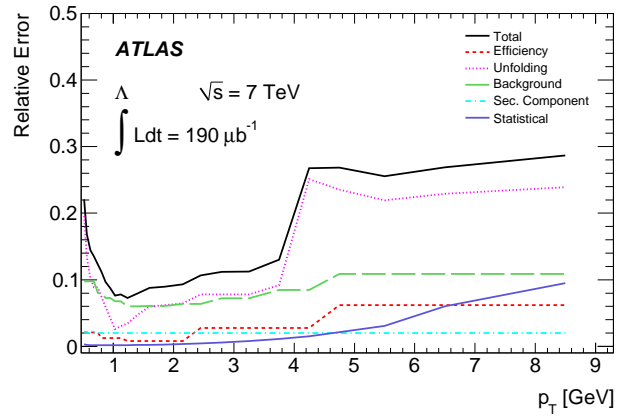}
  \includegraphics[width=0.5\textwidth]{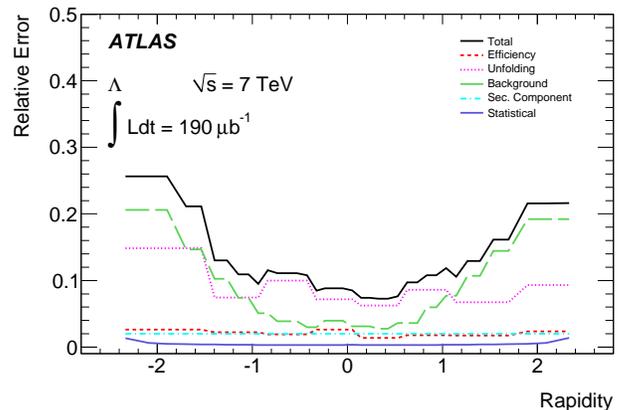}
 \vspace{-1.cm}
  \caption{ The systematic, statistical, and total uncertainties
  versus \pt (top) and rapidity (bottom) of the \LambdaZ candidate in
  $7$~\tev data.
    \label{fig:AllSystematicsLambdas}}
\end{center}
\end{figure}

All the systematic and statistical uncertainties on the \Kshort
distributions in $7$~\tev data are summarized in
Fig.~\ref{fig:AllSystematicsKshorts}.  The total uncertainty, which is
dominated by the systematic component, is at the $5\%$ level in the
peak of the \pt distribution and rises to $10\%$ at higher \pt.  In
the rapidity distribution, the uncertainty is at $4\%$ in the central
region and rises to $6-8\%$ in the forward region.
Figure~\ref{fig:AllSystematicsLambdas} summarizes the systematic and
statistical uncertainties on the \LambdaZ distributions, which are
larger everywhere but show qualitatively similar behavior.

\subsection{\label{subsecLambdaRatioSyst}Systematic uncertainty on the \LambdaZbar/\LambdaZ ratio}

Several systematic effects on the \LambdaZbar/\LambdaZ production
ratio are considered:

\begin{itemize}

\item
The modeling of the interaction cross section for antiprotons in
detector material and its difference from the corresponding cross
section for protons;
\item
The interactions of \LambdaZ and \LambdaZbar baryons in the detector
material before decaying;
\item
Contamination from secondary \LambdaZ and \LambdaZbar baryons.

\end{itemize}

\subsubsection{\label{subsubsecAntiprotonModeling}Modeling of proton and antiproton reconstruction}

The cross sections used by the GEANT4 simulation to model the nuclear
interactions of antiprotons with material have been found to be
over-estimated by the ALICE experiment~\cite{ALICE_Strangeness_Paper,
ALICE_Proton_Paper}.  Any such overestimate biases the correction to
the \LambdaZbar/\LambdaZ ratio described in
Section~\ref{subsecLambdaCorr}. To constrain the accuracy of the
GEANT4 model, patterns of hits on tracks in the outermost two layers
of the SCT are compared between data and MC.  For tracks that have
hits in the three Pixel layers and the first two SCT layers, the
fraction that do not have hits in the outer two layers is a measure of
the inefficiency due to material interactions in those layers.  This
inefficiency is compared between data and MC for protons (antiprotons)
coming from the selected \LambdaZ (\LambdaZbar) candidates and
corrected for background contributions using the invariant-mass
sidebands.  While the data and MC are consistent for proton tracks,
the efficiency for antiprotons is significantly lower in MC than in
data, consistent with the expectation that the interaction cross
section for antiprotons is overestimated in GEANT4.  Comparing the
ratio of antiproton-to-proton efficiency in the outer two layers
between data and MC, a multiplicative correction factor to the
\LambdaZbar/\LambdaZ ratio is extracted as a function of \pt of the
\LambdaZ candidate.  This factor ranges from 0.9 at $\pt=500$~\mev to
0.99 at $\pt=2$~\gev.  (\LambdaZ candidates below $500$~\mev are
rejected as not enough proton candidates are reconstructed at low \pt
to reliably evaluate the correction factor for these candidates.)  As
several correction factors can be formed from various combinations of
hit patterns in the outer two layers, the largest variation among them
is taken as a systematic uncertainty on this correction.  This
uncertainty ranges from $5\%$ at $\pt=500$~\mev to about $1\%$ at
$\pt=2$~\gev.  As an additional cross-check, a sample of protons is
selected using the specific energy loss d$E$/d$x$ measurement in the
Pixel detector~\cite{dEdxNote} and similar data-MC correction factors
are calculated using the efficiency to extend the Pixel tracks to the
SCT.  The results of the d$E$/d$x$ method are consistent with the
hit-pattern study.

\subsubsection{\label{subsubsecLambdaInteraction}Interactions with material before decay and secondary \LambdaZ production}

When evaluated versus the radial position of the decay vertex, the
reconstructed \LambdaZbar/\LambdaZ ratio shows sharp discrete changes
of up to $10\%$ at the detector layers.  In the MC sample, the
dominant cause of this effect is the asymmetric interaction of
\LambdaZ and \LambdaZbar baryons with the detector material before
decay, since such interactions preclude the reconstruction of the
final state of interest.  In addition, roughly $15\%$ of the effect is
caused by secondary baryons asymmetrically produced at the detector
layers by nuclear interactions of other particles.  To constrain the
modeling of these effects in the MC sample, the difference between
data and MC in the change of the ratio at the detector layers is
evaluated.  The data/MC differences at every layer of the tracker are
added together and the sum is assessed as a systematic uncertainty.
Although the value varies in different regions of the detector due to
detector geometry, the largest value of $2.6\%$ (obtained in the
central region) is conservatively assigned to the entire measured
tracking acceptance.  Other evaluations of possible effects of
interactions with material in the MC sample yield an additional
$1.5\%$ uncertainty, for a total uncertainty of $3\%$.  Although the
radial study already includes the effect of secondary \LambdaZ baryons
produced at the detector layers, an additional uncertainty of $1.5\%$
evaluated from the MC sample is assessed to account for the effect of
\LambdaZ baryons produced in the decay of heavier strange baryons.

\subsubsection{\label{subsubsecTotalLambdaSyst}Total uncertainty on \LambdaZ production ratio}

The systematic uncertainties are summarized in
Table~\ref{tab:AllSystematicsLambdaRatio}.  The uncertainty is largest
at low \pt, where it is at the $4.5\%$ level, and approaches the
$3.5\%$ level at higher \pt, where the effect of the proton and
antiproton modeling in GEANT4 is smallest.

\begin{table}[!tb]
\begin{center}
\begin{tabular}{|c|c|}
\hline & Systematic uncertainty \\ \hline \hline Antiproton
 cross section & $\pm$ 1.0-2.8 \%  \\
 \rm{(\pt-dependent)} & \\
 \hline
 Interaction with material & $\pm$ 3.0 \% \\ \hline Secondary
 production & $\pm$ 1.5 \% \\ \hline \hline Total & $\pm$ 3.5-4.4 \%
 \\ \hline
\end{tabular}
\end{center}
\caption{Summary of all systematic uncertainties on the
\LambdaZbar/\LambdaZ production ratio, in $\%$.}
\label{tab:AllSystematicsLambdaRatio}
\end{table}

\begin{figure}[!t]
  \includegraphics[width=0.5\textwidth]{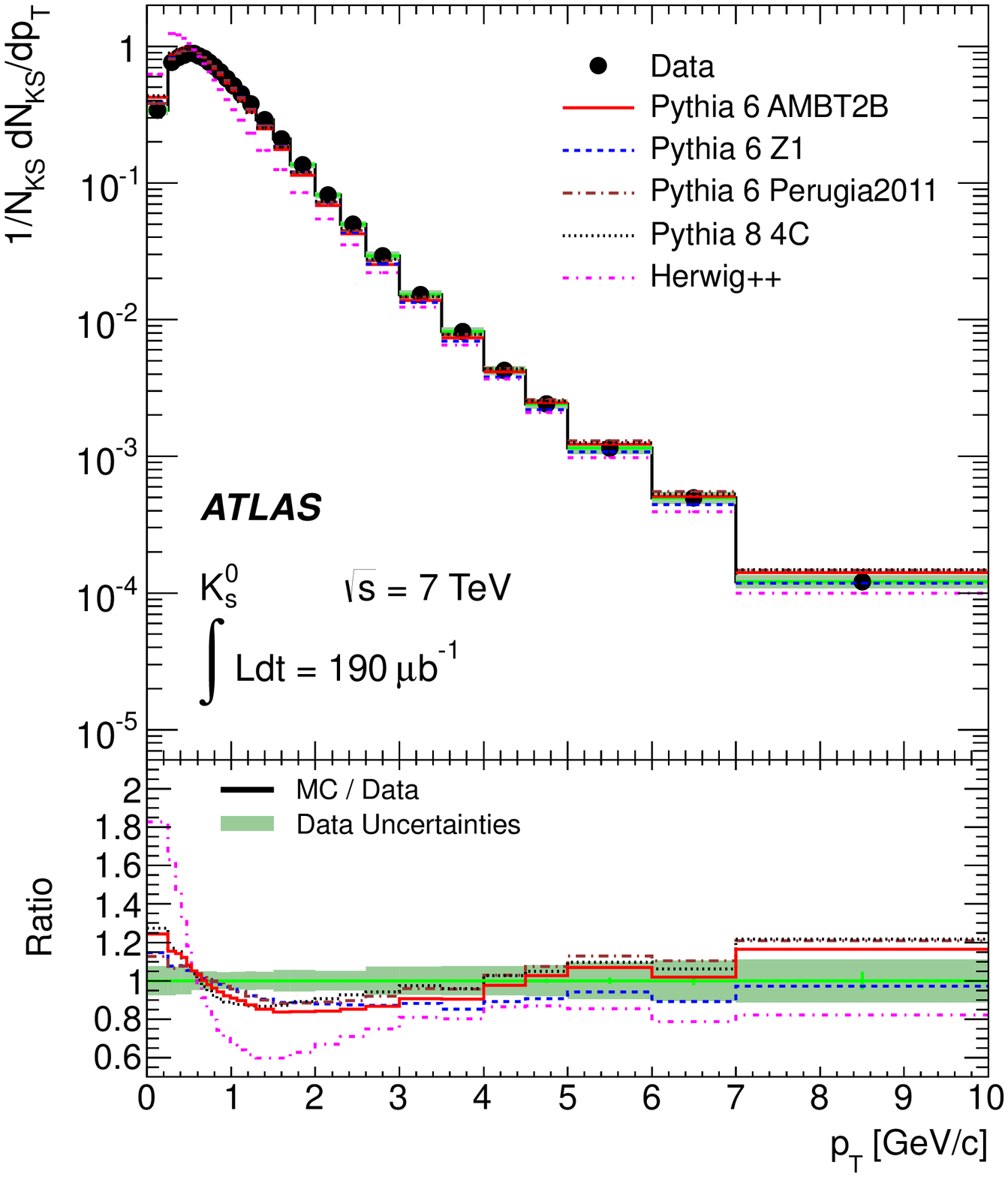}
  \vspace{-0.4cm}
  \caption{The corrected \pt distribution of \Kshort mesons in
  $7$~\tev data compared with the hadron-level distributions in the MC
  samples for a variety of tunes, normalized to unity.  The bottom
  part of the plot shows the ratio of the MC and data distributions,
  with the shaded band showing the statistical and systematic
  uncertainties on the data sample added in
  quadrature.\label{fig:UnfoldingDataPtFinal}}
\end{figure}

\begin{figure}[!h]
  \includegraphics[width=0.5\textwidth]{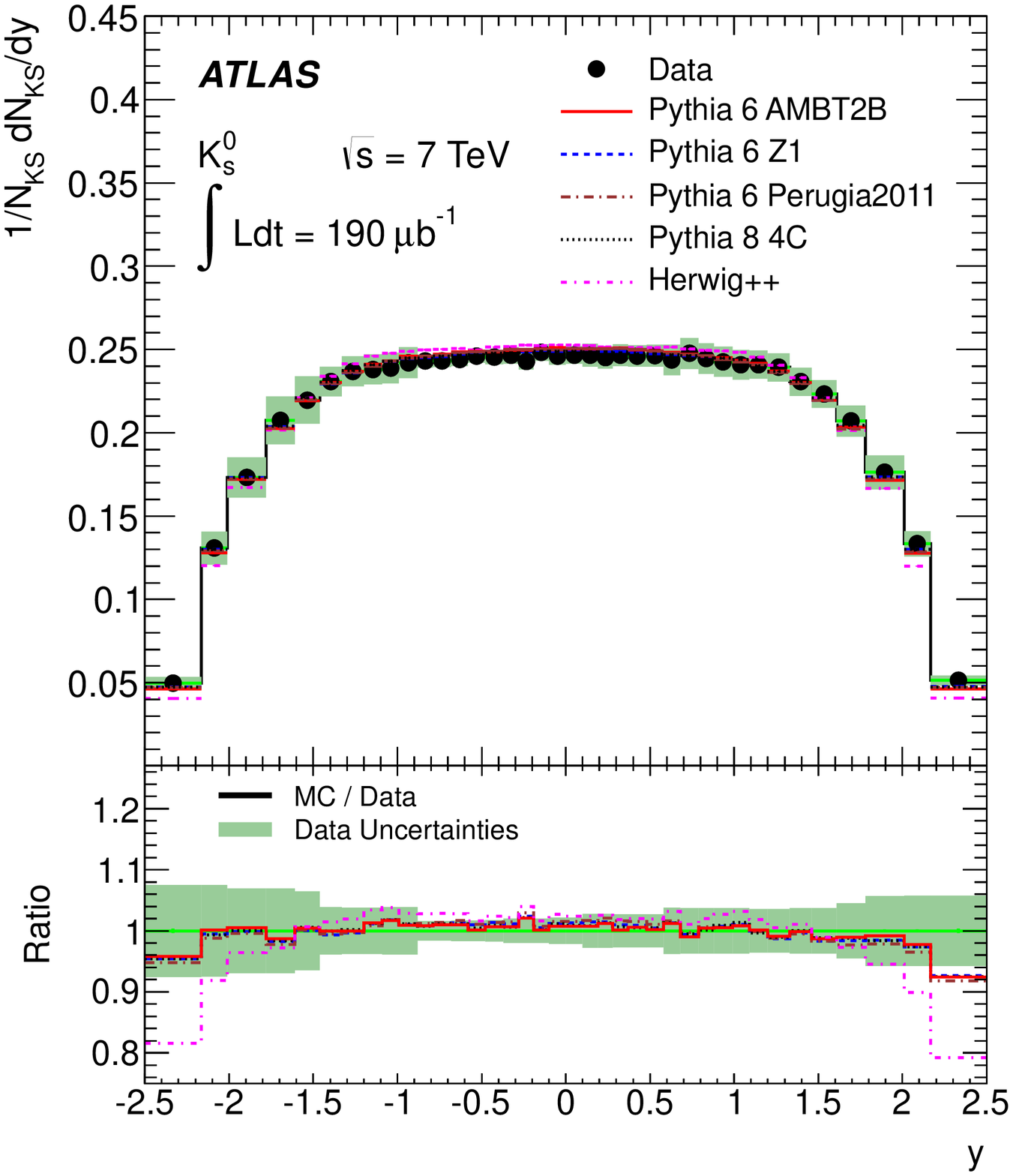}
  \vspace{-0.5cm}
  \caption{The corrected rapidity distribution of \Kshort mesons in
  $7$~\tev data compared with the hadron-level distributions in the MC
  samples for a variety of tunes, normalized to unity.  The bottom
  part of the plot shows the ratio of the MC and data distributions,
  with the shaded band showing the statistical and systematic
  uncertainties on the data sample added in
  quadrature.\label{fig:UnfoldingDataYFinal}}
\end{figure}

\begin{figure}[!h]
  \includegraphics[width=0.5\textwidth]{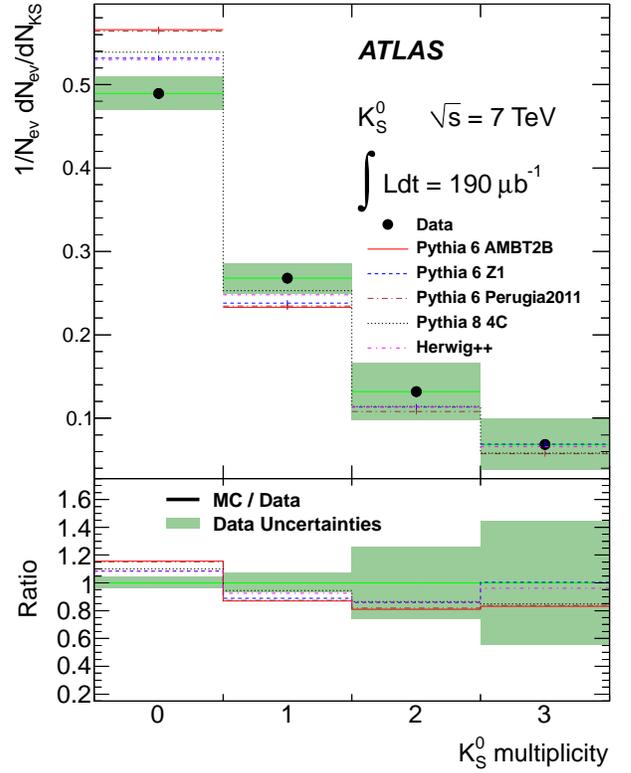}
  \vspace{-0.5cm}
  \caption{The corrected multiplicity distribution of \Kshort mesons
  in $7$~\tev data compared with the hadron-level distributions in the
  MC samples for a variety of tunes, normalized to unity.  The bottom
  part of the plot shows the ratio of the MC and data distributions,
  with the shaded band showing the statistical and systematic
  uncertainties on the data sample added in
  quadrature.\label{fig:UnfoldingDataMultFinal}}
\end{figure}

\section{\label{secResults}Results}

In all corrected distributions, \Kshort mesons are required to have a
flight distance between $4$~mm and $450$~mm and to decay to two
charged pions with $|\eta| < 2.5$ and \pt$>100$~\mev, while \LambdaZ
and \LambdaZbar baryons are required to have $\pt>500~\mev$, flight
distance between $17$~mm and $450$~mm, and to decay to a proton and a
pion with $|\eta| < 2.5$ and \pt$>100$~\mev.  Only \Kshort and
\LambdaZ hadrons consistent with originating from the primary vertex
are considered.  The \pt and rapidity distributions are normalized to
the number of \Kshort or \LambdaZ hadrons, while the multiplicity
distributions are normalized to the total number of events with two
charged particles satisfying $\pt>100~\mev$ and $|\eta|<2.5$.  The
multiplicity distributions are corrected for branching fractions to
the measured final states using world-average values~\cite{PDGRef}.
Predictions from several MC generators are shown with the same
acceptance requirements.

\begin{figure}[!tbh]
  \includegraphics[width=0.5\textwidth]{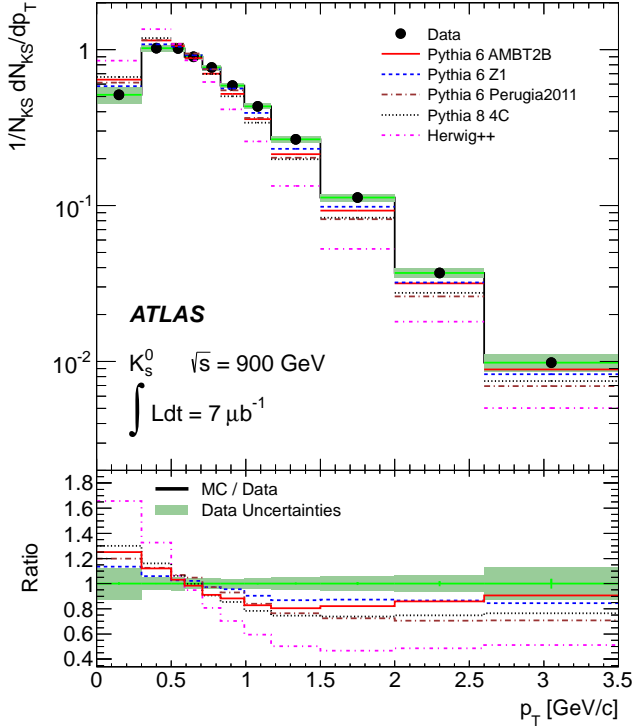}
  \vspace{-0.5cm}
  \caption{The corrected \pt distribution of \Kshort mesons in
  $900$~\gev data compared with the hadron-level distributions in the
  MC samples for a variety of tunes, normalized to unity.  The bottom
  part of the plot shows the ratio of the MC and data distributions,
  with the shaded band showing the statistical and systematic
  uncertainties on the data sample added in
  quadrature.\label{fig:UnfoldingDataPtFinal-900GeV}}
\end{figure}

\begin{figure}[!tbh]
  \includegraphics[width=0.5\textwidth]{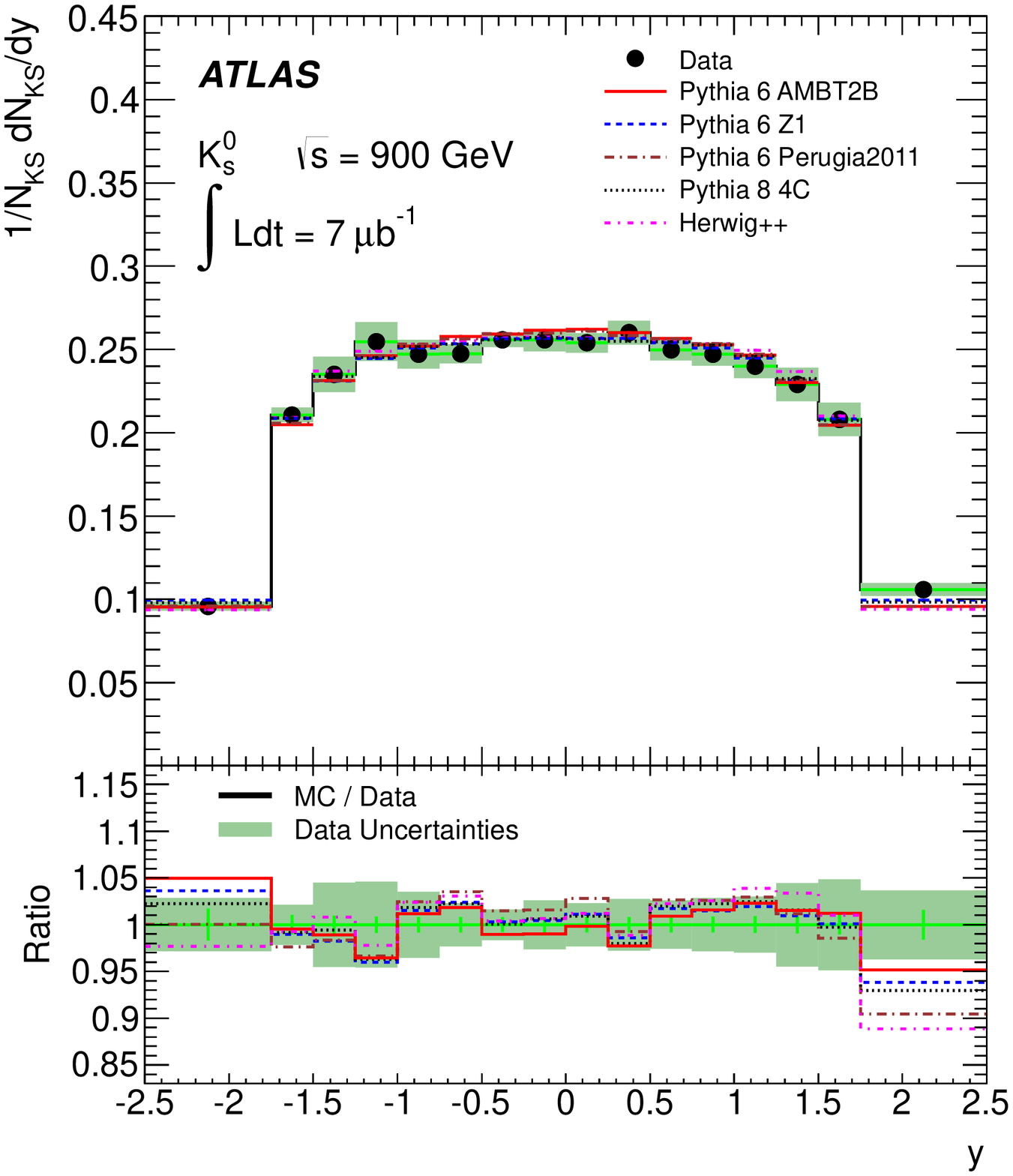}
  \vspace{-0.5cm}
  \caption{The corrected rapidity distribution of \Kshort mesons in
  $900$~\gev data compared with the hadron-level distributions in the
  MC samples for a variety of tunes, normalized to unity.  The bottom
  part of the plot shows the ratio of the MC and data distributions,
  with the shaded band showing the statistical and systematic
  uncertainties on the data sample added in
  quadrature.\label{fig:UnfoldingDataYFinal-900GeV}}
\end{figure}

\begin{figure}[!tbh]
  \includegraphics[width=0.5\textwidth]{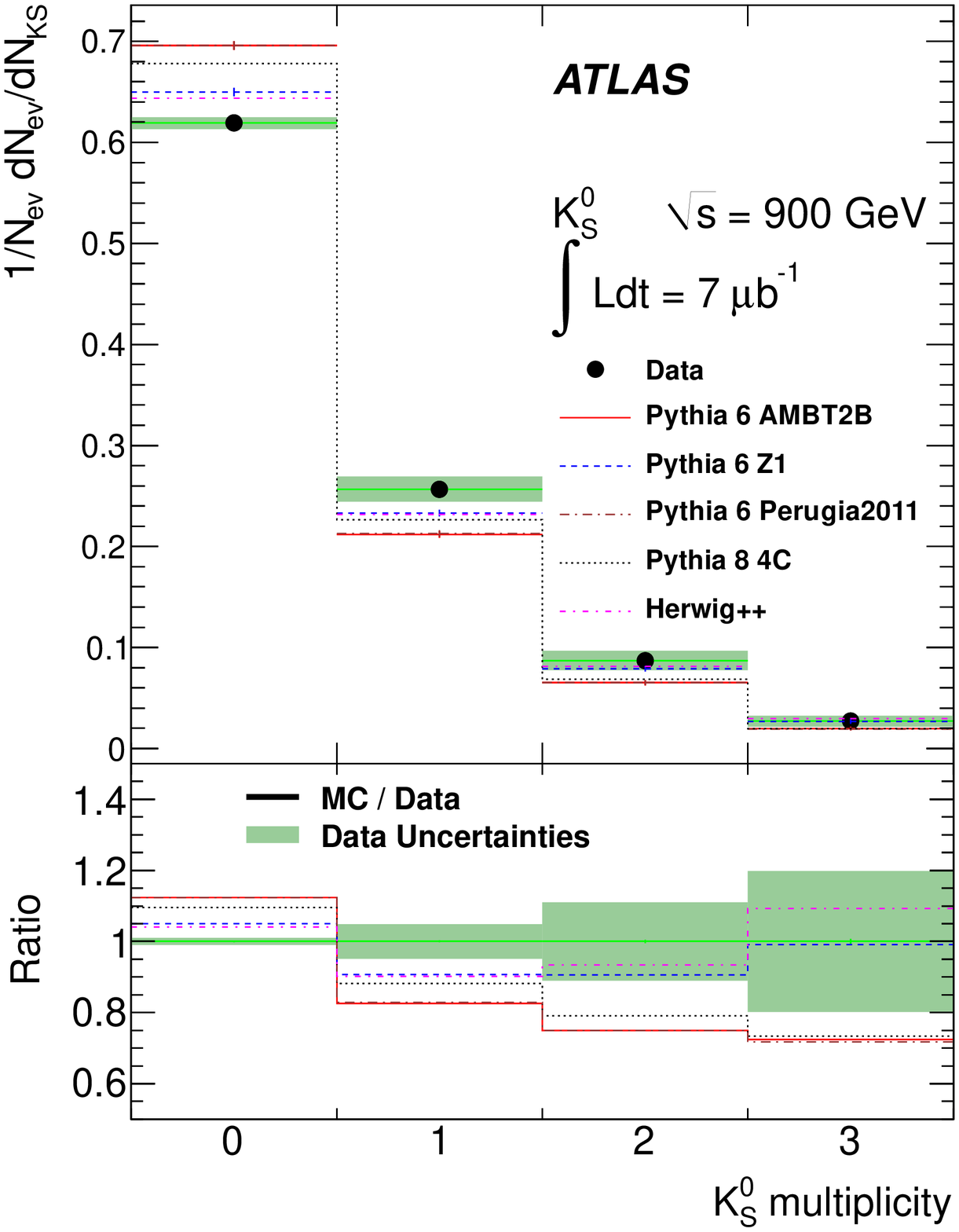}
  \vspace{-0.5cm}
  \caption{The corrected multiplicity distribution of \Kshort mesons
  in $900$~\gev data compared with the hadron-level distributions in
  the MC samples for a variety of tunes, normalized to unity.  The
  bottom part of the plot shows the ratio of the MC and data
  distributions, with the shaded band showing the statistical and
  systematic uncertainties on the data sample added in
  quadrature.\label{fig:UnfoldingDataMultFinal-900GeV}}
\end{figure}

\begin{figure}[!tbh]
  \includegraphics[width=0.5\textwidth]{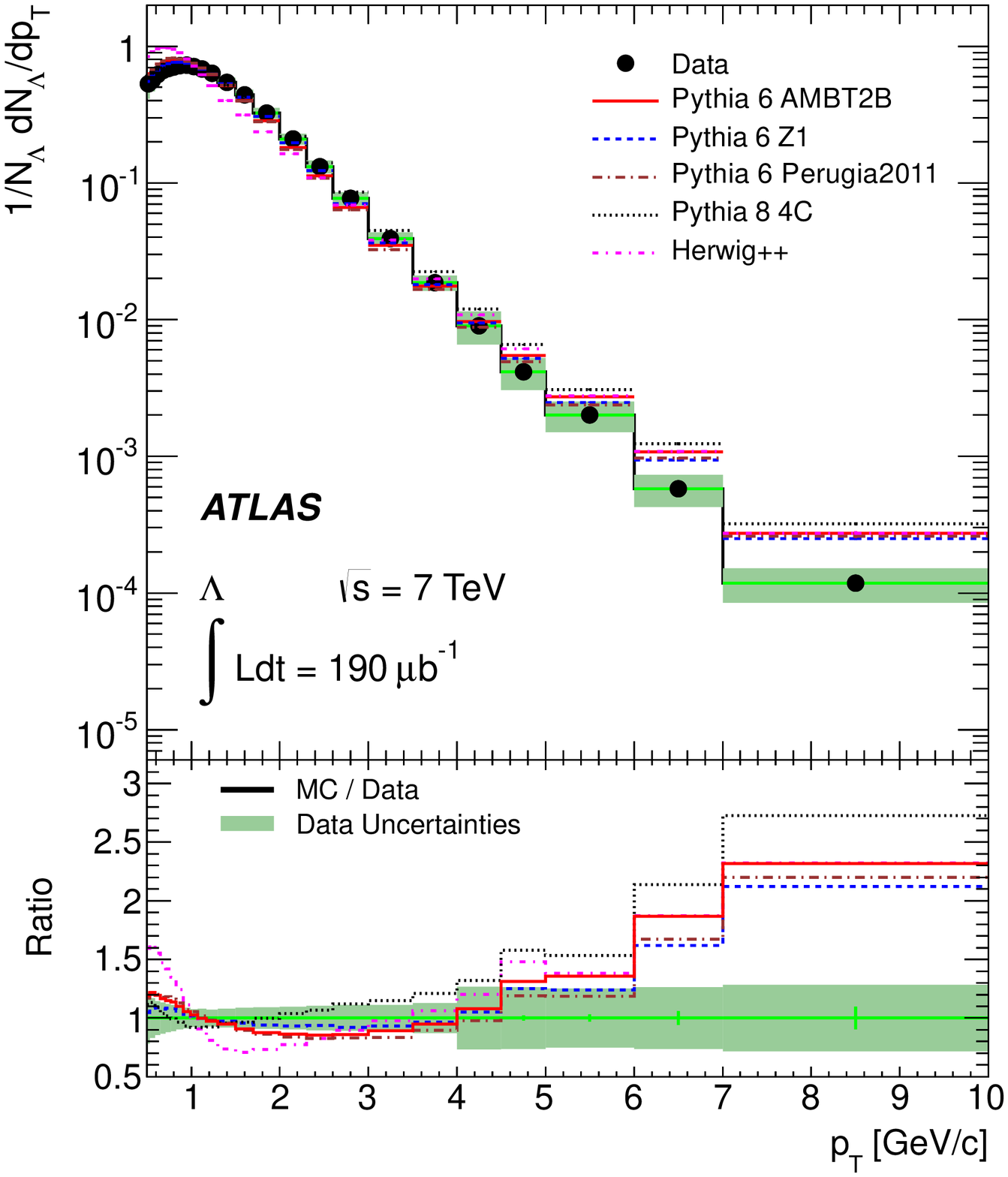}
  \vspace{-0.5cm}
  \caption{The corrected \pt distribution of \LambdaZ baryons in
  $7$~\tev data compared with the hadron-level distributions in the MC
  samples for a variety of tunes, normalized to unity.  The bottom
  part of the plot shows the ratio of the MC and data distributions,
  with the shaded band showing the statistical and systematic
  uncertainties on the data sample added in
  quadrature.\label{fig:UnfoldingDataPtFinal-Lambdas}}
\end{figure}

\begin{figure}[!tbh]
  \includegraphics[width=0.5\textwidth]{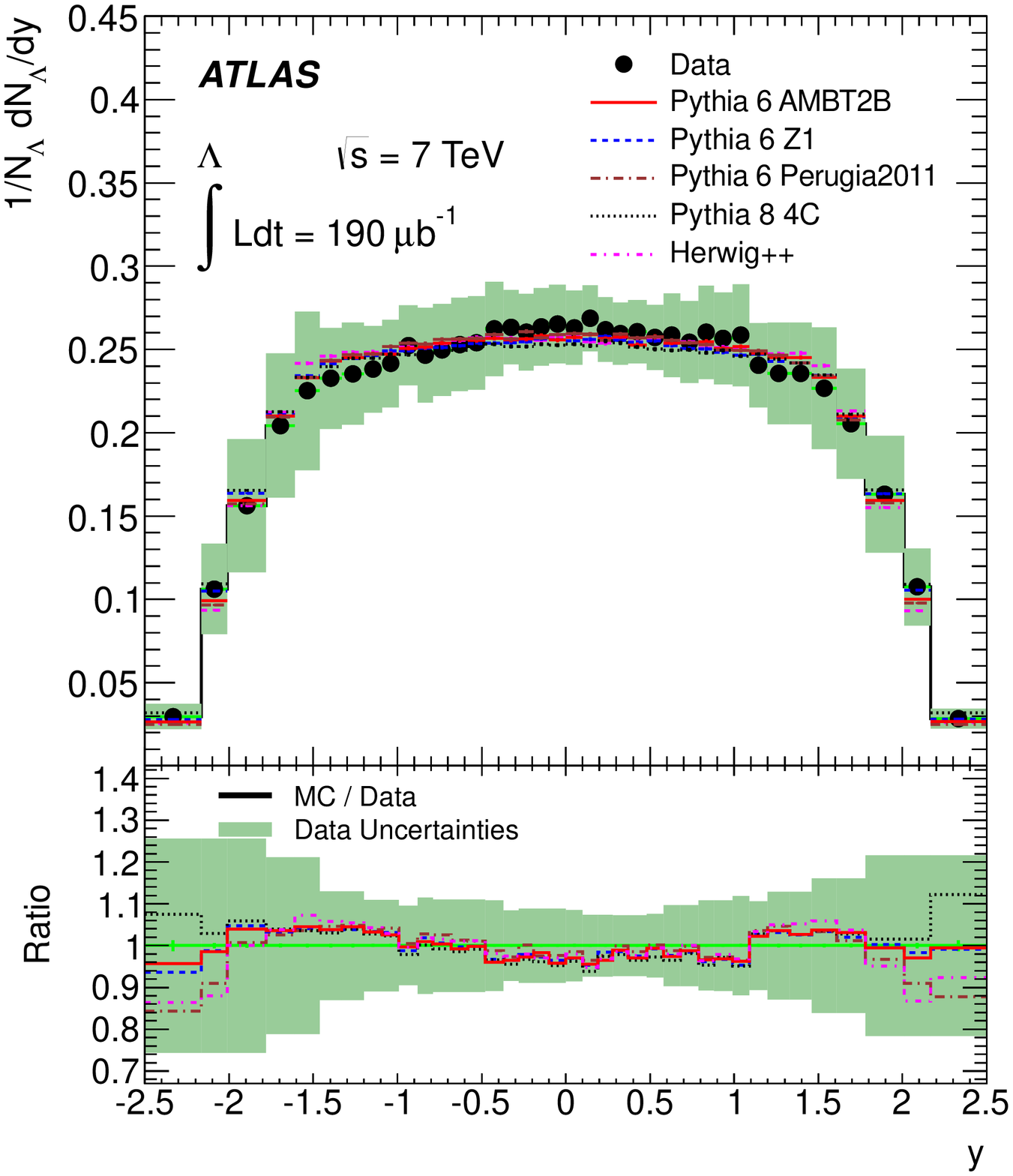}
  \vspace{-0.5cm}
  \caption{The corrected rapidity distribution of \LambdaZ baryons in
  $7$~\tev data compared with the hadron-level distributions in the MC
  samples for a variety of tunes, normalized to unity.  The bottom
  part of the plot shows the ratio of the MC and data distributions,
  with the shaded band showing the statistical and systematic
  uncertainties on the data sample added in
  quadrature.\label{fig:UnfoldingDataYFinal-Lambdas}}
\end{figure}

\begin{figure}[!b]
  \includegraphics[width=0.5\textwidth]{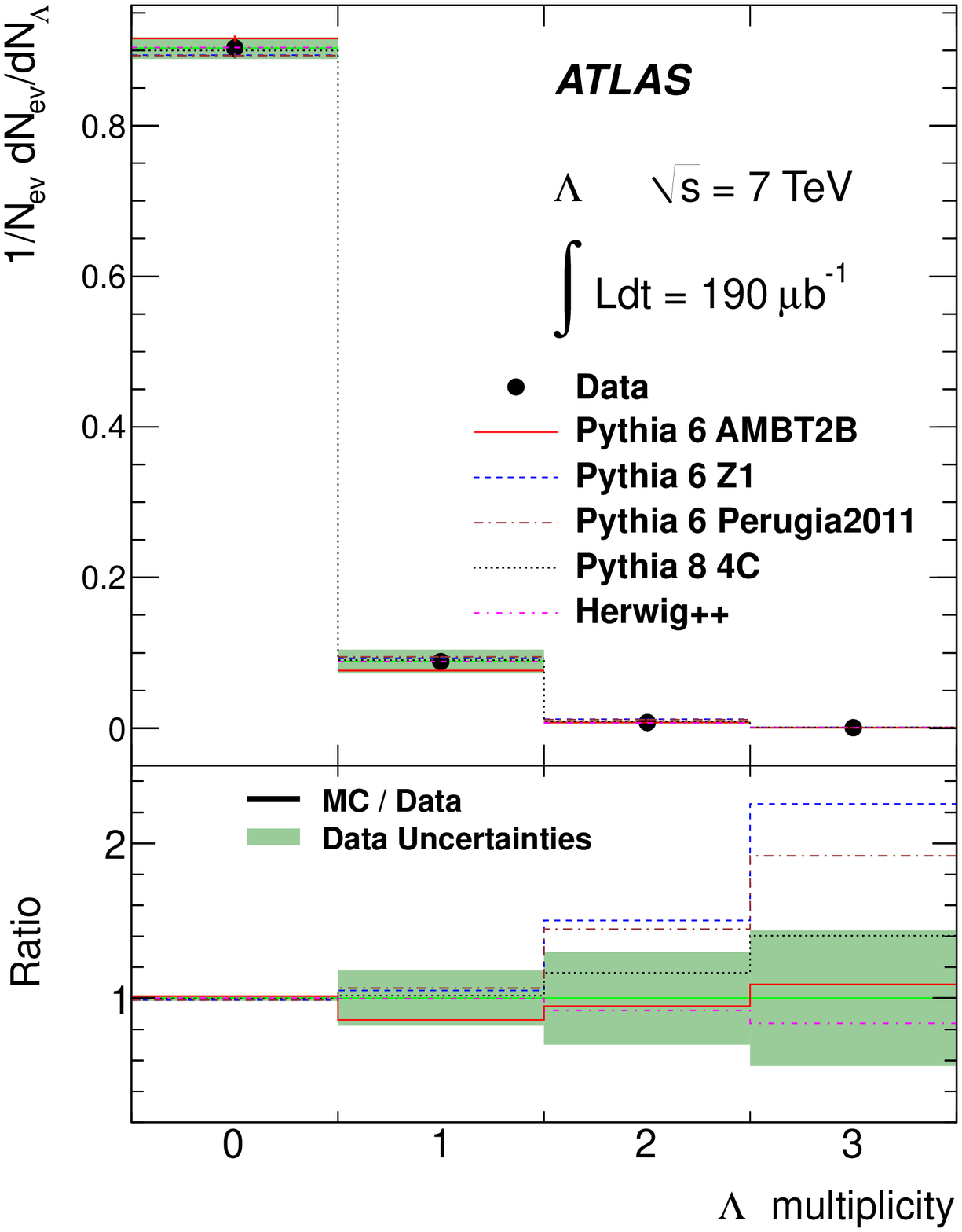}
  \vspace{-0.5cm}
  \caption{The corrected multiplicity distribution of \LambdaZ baryons
  in $7$~\tev data compared with the hadron-level distributions in the
  MC samples for a variety of tunes, which are normalized to unity.
  The bottom part of the plot shows the ratio of the MC and data
  distributions, with the shaded band showing the statistical and
  systematic uncertainties on the data sample added in
  quadrature.\label{fig:UnfoldingDataMultFinalLambdas}}
\end{figure}

\begin{figure}[!tbh]
  \includegraphics[width=0.5\textwidth]{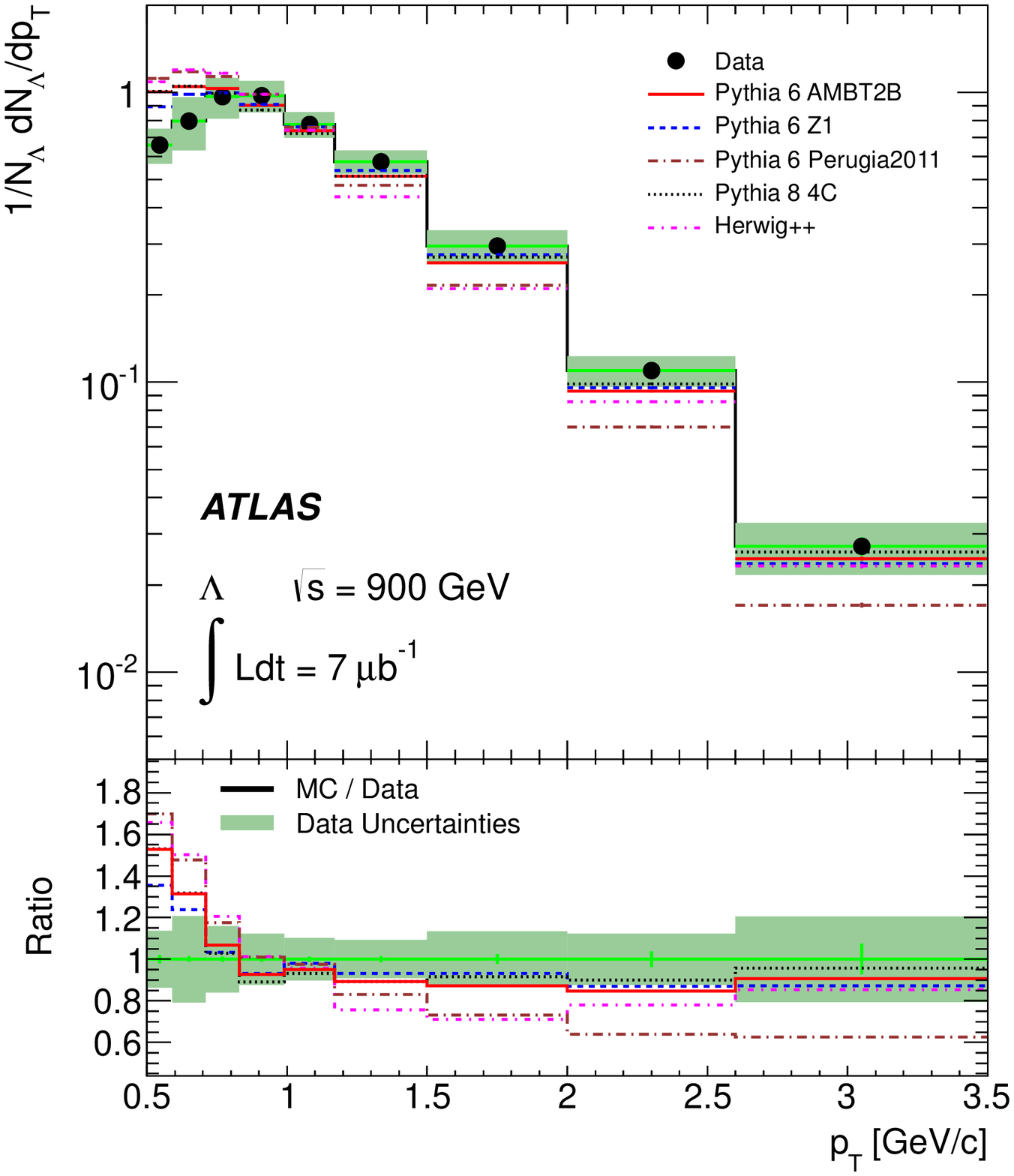}
  \vspace{-0.5cm}
  \caption{The corrected \pt distribution of \LambdaZ baryons in
  $900$~\gev data compared with the hadron-level distributions in the
  MC samples for a variety of tunes, normalized to unity.  The bottom
  part of the plot shows the ratio of the MC and data distributions,
  with the shaded band showing the statistical and systematic
  uncertainties on the data sample added in
  quadrature.\label{fig:UnfoldingDataPtFinal-900GeV-Lambdas}}
\end{figure}

\begin{figure}[!tbh]
  \includegraphics[width=0.5\textwidth]{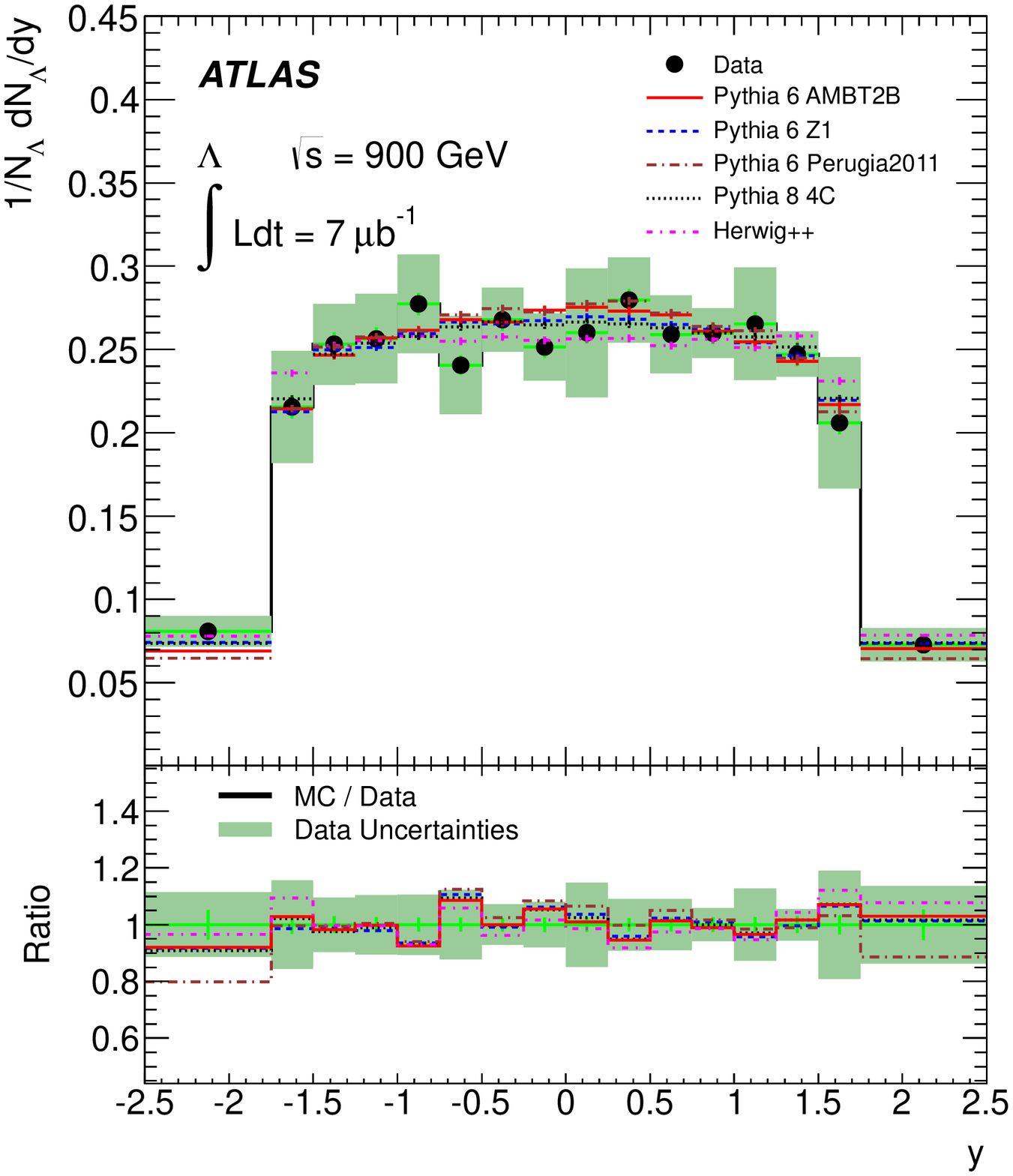}
  \vspace{-0.5cm}
  \caption{The corrected rapidity distribution of \LambdaZ baryons in
  $900$~\gev data compared with the hadron-level distributions in the
  MC samples for a variety of tunes, normalized to unity.  The bottom
  part of the plot shows the ratio of the MC and data distributions,
  with the shaded band showing the statistical and systematic
  uncertainties on the data sample added in
  quadrature.\label{fig:UnfoldingDataYFinal-900GeV-Lambdas}}
\end{figure}

\begin{figure}[!b]
  \includegraphics[width=0.5\textwidth]{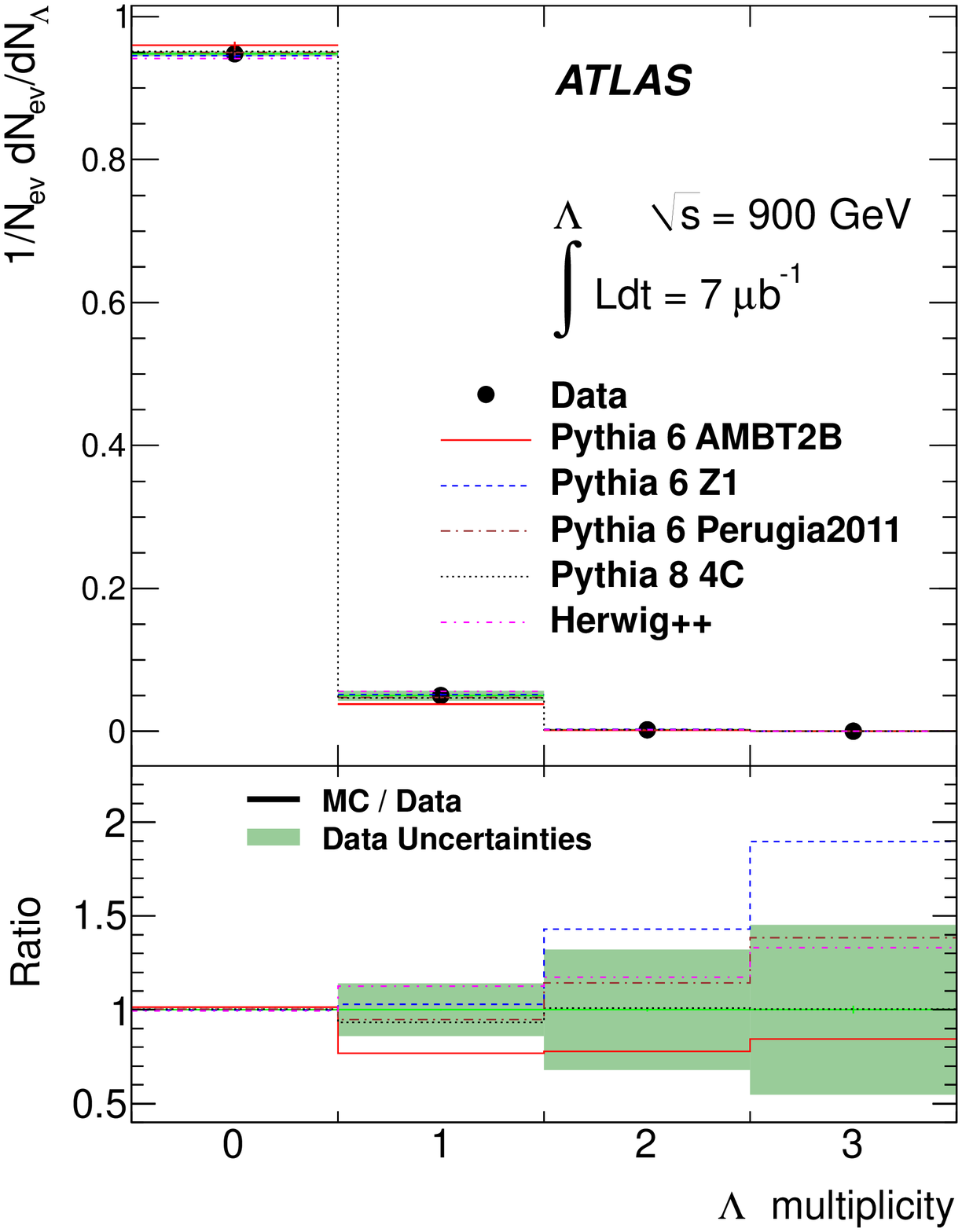}
  \vspace{-0.5cm}
  \caption{The corrected multiplicity distribution of \LambdaZ baryons
  in $900$~\gev data compared with the hadron-level distributions in
  the MC samples for a variety of tunes, which are normalized to
  unity.  The bottom part of the plot shows the ratio of the MC and
  data distributions, with the shaded band showing the statistical and
  systematic uncertainties on the data sample added in
  quadrature.\label{fig:UnfoldingDataMultFinalLambdas-900GeV}}
\end{figure}

Figures~\ref{fig:UnfoldingDataPtFinal}
and~\ref{fig:UnfoldingDataYFinal} show the corrected production
distributions of \Kshort mesons versus transverse momentum and
rapidity, respectively, in $7$~\tev data.
Figure~\ref{fig:UnfoldingDataMultFinal} shows the distribution of
\Kshort multiplicity in $7$~\tev data.
Figures~\ref{fig:UnfoldingDataPtFinal-900GeV}
and~\ref{fig:UnfoldingDataYFinal-900GeV} show the corrected production
distributions of \Kshort mesons versus transverse momentum and
rapidity, respectively, in $900$~\gev data, while
Fig.~\ref{fig:UnfoldingDataMultFinal-900GeV} shows the distribution of
\Kshort multiplicity in $900$~\gev data.
Figures~\ref{fig:UnfoldingDataPtFinal-Lambdas}
and~\ref{fig:UnfoldingDataYFinal-Lambdas} show the corrected
production distributions of \LambdaZ baryons versus transverse
momentum and rapidity, respectively, in $7$~\tev data, while
Fig.~\ref{fig:UnfoldingDataMultFinalLambdas} shows the distribution of
\LambdaZ multiplicity in $7$~\tev data.
Figures~\ref{fig:UnfoldingDataPtFinal-900GeV-Lambdas}
and~\ref{fig:UnfoldingDataYFinal-900GeV-Lambdas} show the corrected
production distributions of \LambdaZ baryons versus transverse
momentum and rapidity, respectively, in $900$~\gev data, while
Fig.~\ref{fig:UnfoldingDataMultFinalLambdas-900GeV} shows the
distribution of \LambdaZ multiplicity in $900$~\gev data.

\begin{figure}[!b]
  \includegraphics[width=0.5\textwidth]{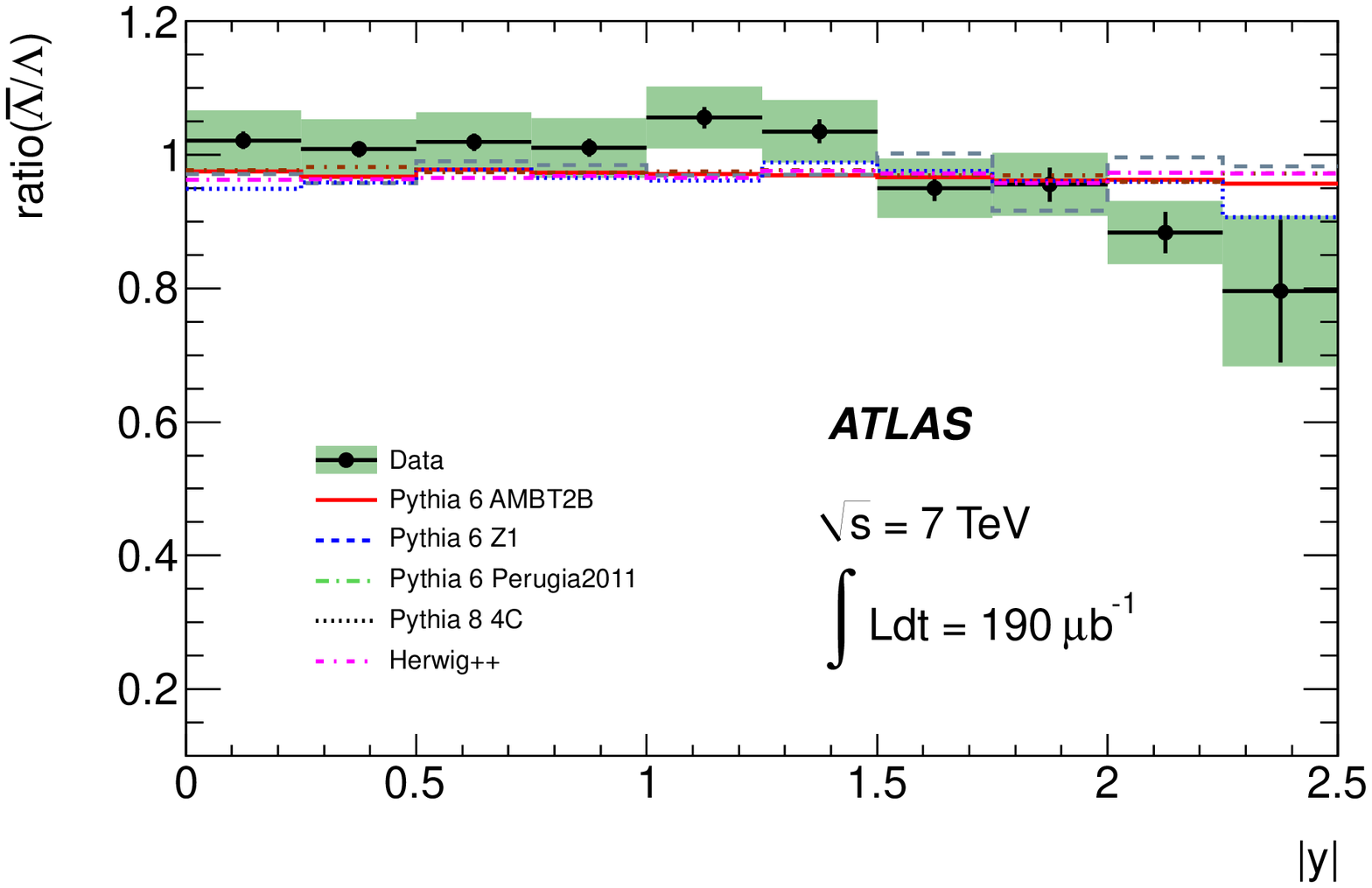}   
  \vspace{-0.5cm}
  \caption{The production ratio between \LambdaZbar and \LambdaZ
  baryons in $7$~\tev data versus the absolute value of the rapidity.
  The error bars show the statistical uncertainties while the band
  shows statistical and systematic uncertainties added in
  quadrature.\label{fig:Lambda-Y-Final}}
\end{figure}

\begin{figure}[!b]
  \includegraphics[width=0.5\textwidth]{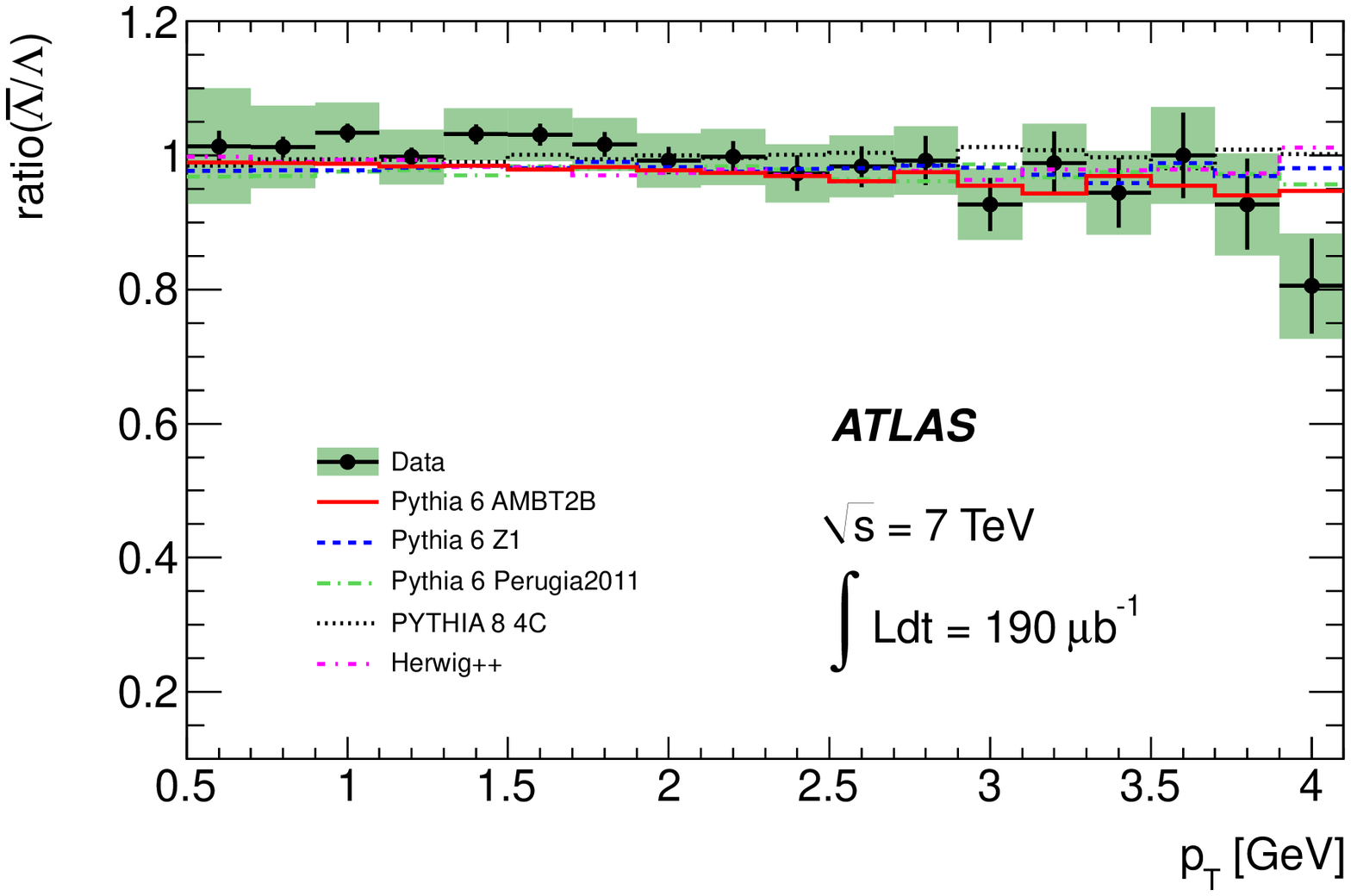}   
  \vspace{-0.5cm}
  \caption{The production ratio between \LambdaZbar and \LambdaZ
  baryons in $7$~\tev data versus \pt.  The error bars show the
  statistical uncertainties while the band shows statistical and
  systematic uncertainties added in
  quadrature.\label{fig:Lambda-Pt-Final}}
\end{figure}

\begin{figure}[!b]
  \includegraphics[width=0.5\textwidth]{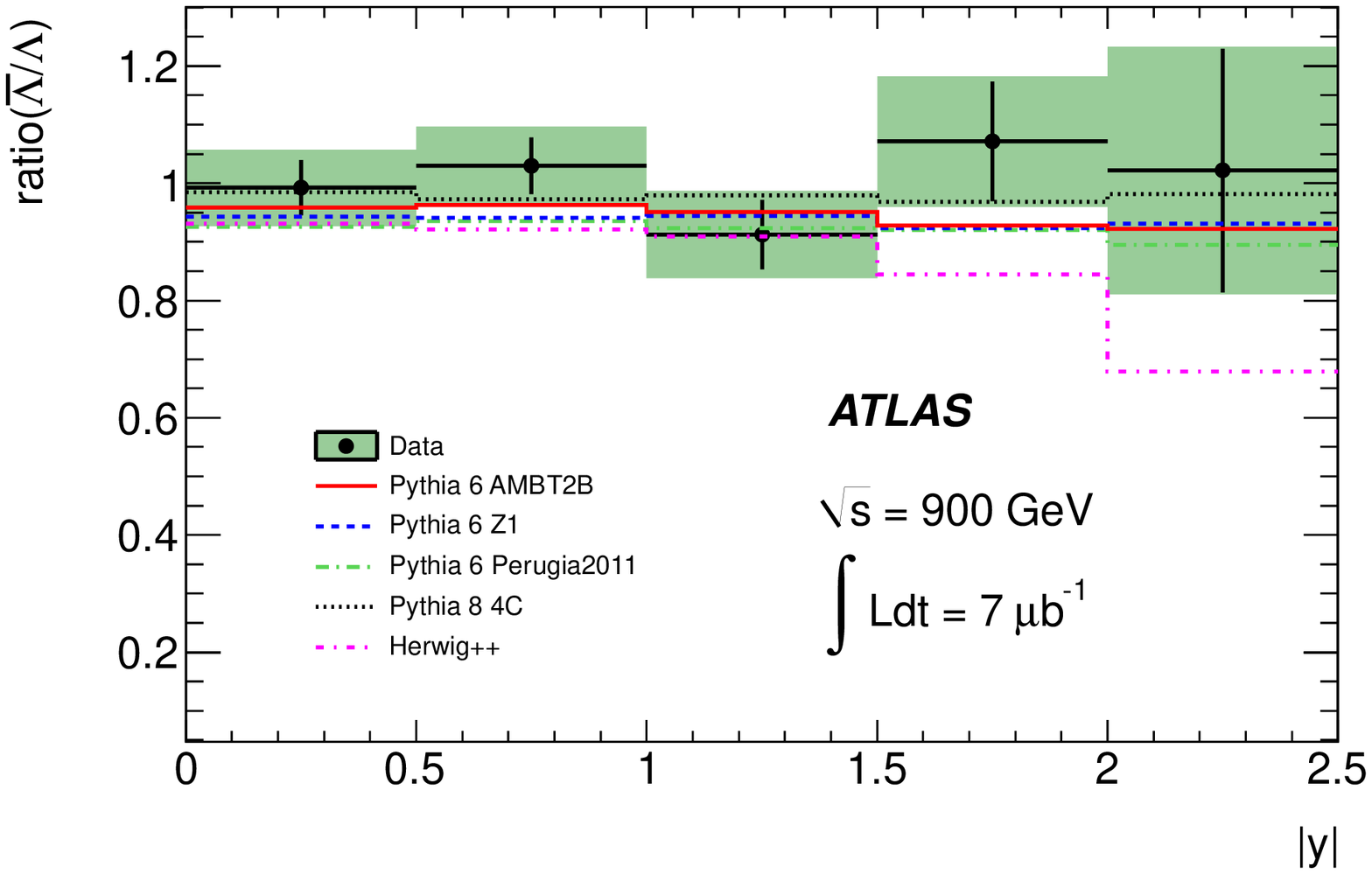}   
  \vspace{-0.5cm}
  \caption{The production ratio between \LambdaZbar and \LambdaZ
  baryons in $900$~\gev data versus the absolute value of the
  rapidity.  The error bars show the statistical uncertainties while
  the band shows statistical and systematic uncertainties added in
  quadrature.\label{fig:Lambda-Y-Final-900GeV}}
\end{figure}

\begin{figure}[!tb]
  \includegraphics[width=0.5\textwidth]{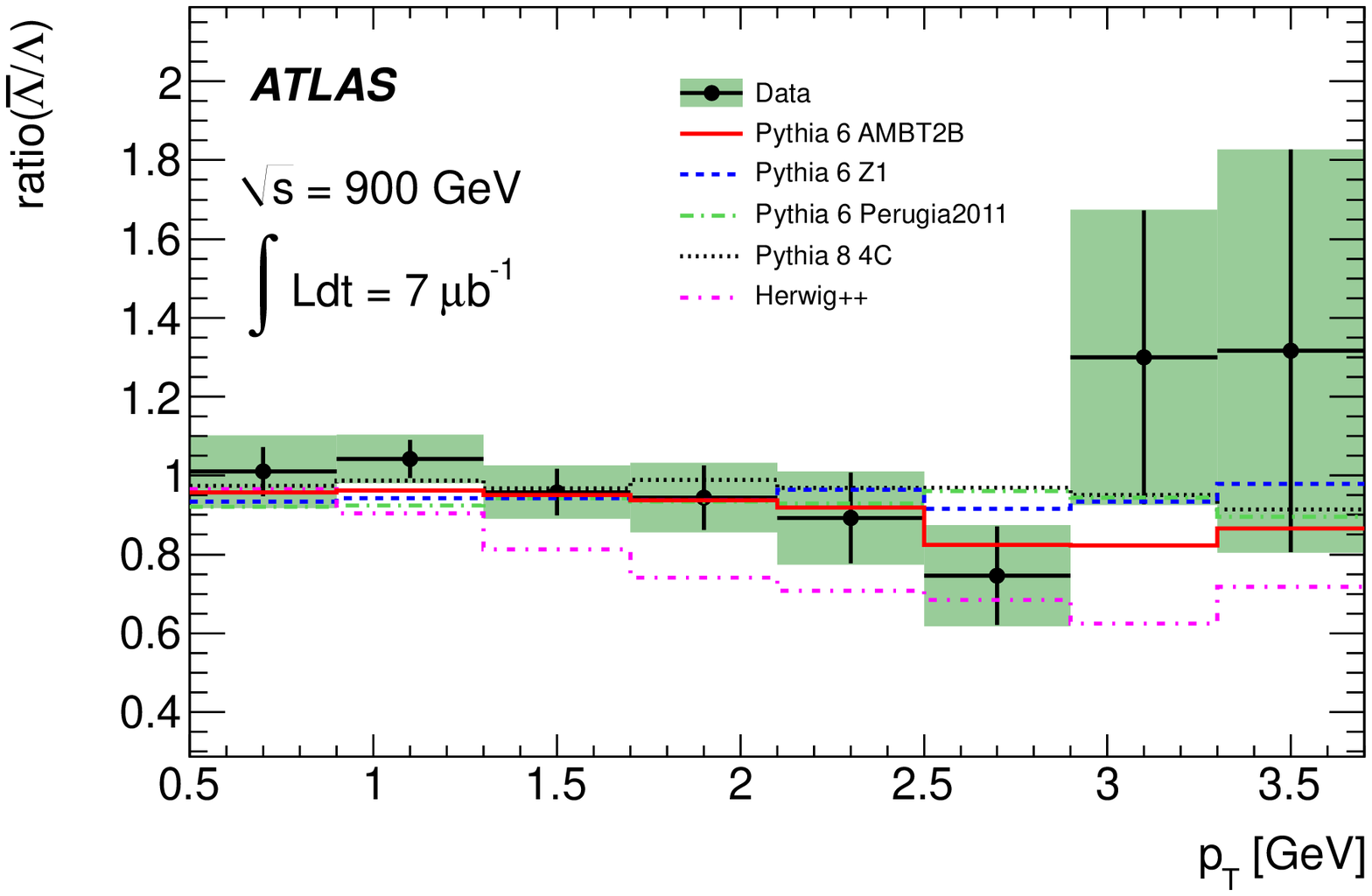}   
  \vspace{-0.5cm}
  \caption{The production ratio between \LambdaZbar and \LambdaZ
  baryons in $900$~\gev data versus \pt.  The error bars show the
  statistical uncertainties while the band shows statistical and
  systematic uncertainties added in
  quadrature.\label{fig:Lambda-Pt-Final-900GeV}}
\end{figure}

The fully corrected \LambdaZbar/\LambdaZ production ratio is shown in
Fig.~\ref{fig:Lambda-Y-Final} versus the absolute value of rapidity
and in Fig.~\ref{fig:Lambda-Pt-Final} versus \pt, along with
predictions from several MC models.  The ratio is shown only for
candidates with $\pt>500$~\mev.  The corrected ratio is consistent
with unity everywhere, while the uncertainties within the barrel,
transition, and endcap regions in rapidity are highly correlated due
to common detector corrections and systematic effects.  The
measurement is statistically limited at higher \pt, while at lower \pt
the systematic effects of the modeling of antiproton reconstruction in
simulation dominate the uncertainty.
Figs.~\ref{fig:Lambda-Y-Final-900GeV}
and~\ref{fig:Lambda-Pt-Final-900GeV} show the \LambdaZbar/\LambdaZ
production ratio in $900$~\gev data.\\ \\

\section{\label{secDiscussion}Discussion and Conclusions}

While the shape of the rapidity distribution for \Kshort mesons in
$7$~\tev data agrees with the hadron-level \textsc{pythia}
distributions to $5\%$ (Fig.~\ref{fig:UnfoldingDataYFinal}), the
\textsc{pythia} tunes fall more slowly than data versus \pt above
$2$~\gev (Fig.~\ref{fig:UnfoldingDataPtFinal}), although the
deviations are within $15\%$ everywhere except at the lowest \pt bin.
This shape discrepancy is much improved from the earlier generation of
tunes used in ATLAS, as the current models have been tuned using
minimum-bias data from the LHC experiments. The best agreement is
observed in the \textsc{pythia}6 Z1 tune, but the variation among the
\textsc{pythia} tunes is small.  Although the shape of the
\textsc{herwig++} distribution (UE7-2 tune) agrees with data above
$3$~\gev, it does a poor job at lower momenta.  All of the MC models
underestimate the number of \Kshort mesons per minimum-bias event
(Fig.~\ref{fig:UnfoldingDataMultFinal}), but the experimental
uncertainties preclude drawing a significant conclusion about the
shape of the multiplicity distribution.

In the case of \LambdaZ baryons at $7$~\tev, all of the tunes disagree
with data at high-\pt and to a greater degree than in the \Kshort case
(Fig.~\ref{fig:UnfoldingDataPtFinal-Lambdas}).  The worst agreement is
for \textsc{pythia8}, which deviates from data by a factor of about
2.5 at the highest measured momenta.  The Perugia2011 and Z1 tunes
also significantly overestimate the production of \LambdaZ baryons per
event at both energies (Fig.~\ref{fig:UnfoldingDataMultFinalLambdas}).

The AMBT2B tune agrees with $900$~\gev data for \Kshort mesons to
better than about $25\%$ across the whole \pt range
(Fig.~\ref{fig:UnfoldingDataPtFinal-900GeV}), while \textsc{herwig++}
(MU900-2 tune) disagrees with data more strongly than in the $7$~\tev
case (UE7-2 tune).  The number of \Kshort mesons per event
(Fig.~\ref{fig:UnfoldingDataMultFinal-900GeV}) is underestimated as in
the $7$~\tev data.  In the \LambdaZ \pt distribution
(Fig.~\ref{fig:UnfoldingDataPtFinal-900GeV-Lambdas}) all tunes agree
with data better at $900$~\gev than at $7$~\tev.

\begin{figure}[!tb]
  \includegraphics[width=0.5\textwidth]{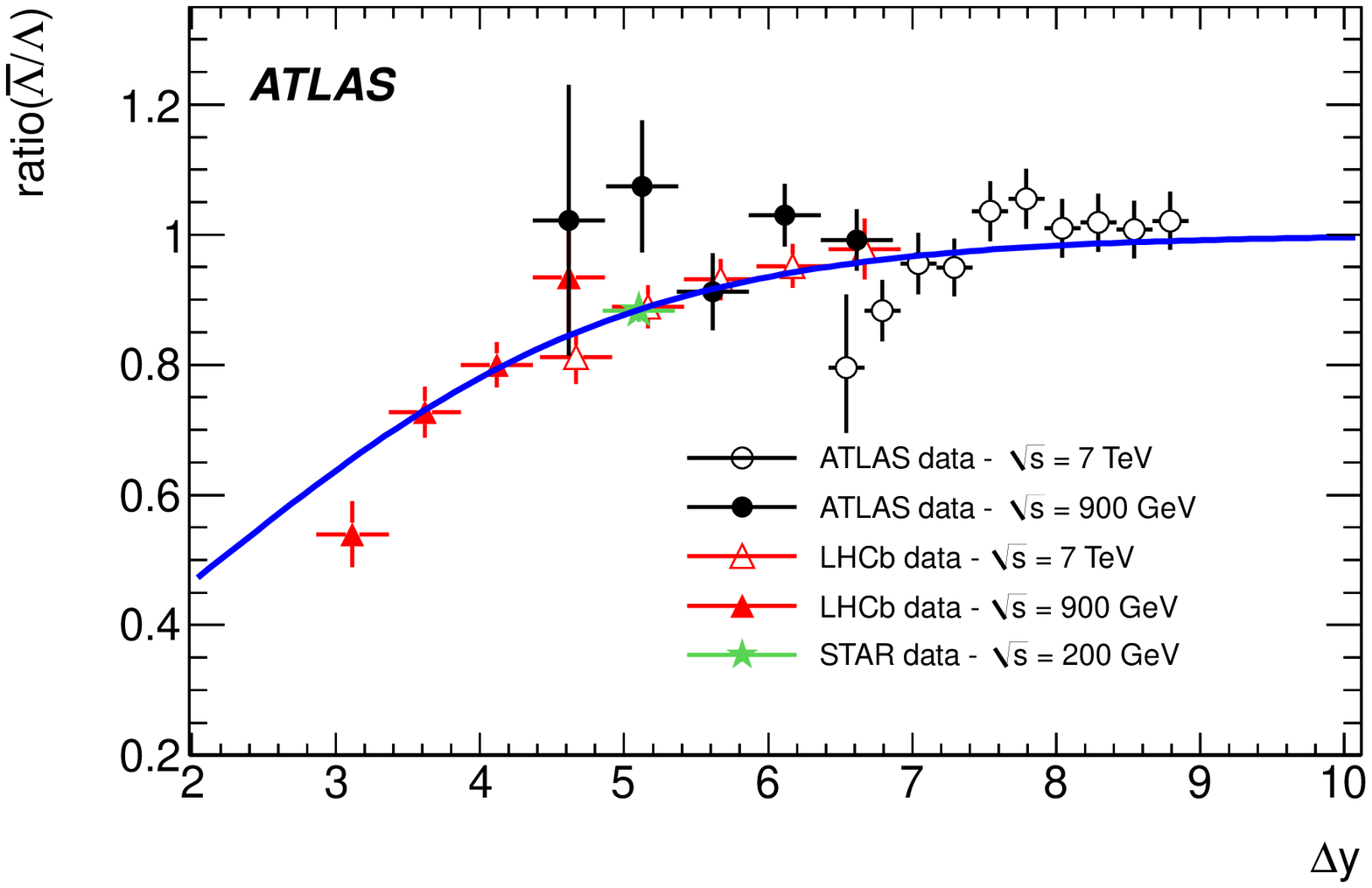}
  \vspace{-0.5cm}
  \caption{The production ratio between \LambdaZbar and \LambdaZ
  baryons measured by ATLAS and other experiments versus the rapidity
  difference with respect to the beam.  The error bars on the ATLAS
  data show statistical and systematic uncertainties added in
  quadrature.  The solid line shows the fit to all data points
  described in the text.\label{fig:ExperimentComparison}}
\end{figure}

The \LambdaZbar/\LambdaZ production ratio at both energies is
consistent with unity everywhere and does not show a significant
variation with either rapidity or \pt within our total uncertainties.
\textsc{herwig++} (MU900-2 tune) shows a decrease in the ratio versus
both \pt and rapidity at $900$~\gev that is not reproduced by the data
(Fig.~\ref{fig:Lambda-Y-Final-900GeV}).  The measurement is consistent
with other antibaryon-baryon ratio measurements from the ALICE, LHCb,
and STAR experiments~\cite{ALICE_Strangeness_Paper,
ALICE_Proton_Paper, LHCb_Paper, STAR_Paper}.  Measurements from
several other experiments are shown in
Fig.~\ref{fig:ExperimentComparison} in terms of the difference between
the rapidity of the observed baryons and the rapidity of the proton
beam ($y_{\rm beam} \approx 8.9$ and $6.9$ at $7$~\tev and $900$~\gev,
respectively), along with a combined fit to the following functional
form~\cite{ALICE_Proton_Paper} that has been found empirically to
describe the data at several energies:
\begin{equation}
\frac{1}{\rm ratio} = 1+C \times e^{(\alpha_J-\alpha_P)\Delta y},
\end{equation}
where $\alpha_J$ and $\alpha_P$ are related to the string-junction and
Pomeron models, respectively.  Following
Ref.~\cite{ALICE_Proton_Paper}, the parameters are fixed to
$\alpha_J=0.5$ and $\alpha_P=1.2$ and the value $C=4.6\pm0.5$ is
obtained from the fit, assuming that the uncertainties are
uncorrelated among the measurements.

In summary, measurements are presented of the \pt, rapidity, and
multiplicity distributions of \Kshort and \LambdaZ production in $pp$
collisions at $\sqrt{s} = 0.9$ and $7$~\tev with the ATLAS detector,
as well as the \LambdaZbar/\LambdaZ production ratio.  The data
results are compared with several recent \textsc{pythia} MC models
that were tuned on early LHC data and are found to describe the data
significantly better than the previous generation of tunes.  All
\textsc{pythia} tunes underestimate the production of \Kshort mesons
per event and overestimate the production of \LambdaZ baryons per
event.  The \textsc{herwig++} tunes significantly disagree with data
in both \pt and multiplicity at the respective energies.  Despite the
general improvement in the agreement with data, no considered model
agrees in both the \pt and multiplicity quantities simultaneously,
indicating the need for further model development.  The
\LambdaZbar/\LambdaZ ratio is consistent with unity in data,
indicating that no significant transport of baryon number to
mid-rapidities is present, in accordance with SM predictions and
measurements from other experiments.

\section{Acknowledgements}

We thank CERN for the very successful operation of the LHC, as well as the
support staff from our institutions without whom ATLAS could not be
operated efficiently.

We acknowledge the support of ANPCyT, Argentina; YerPhI, Armenia; ARC,
Australia; BMWF, Austria; ANAS, Azerbaijan; SSTC, Belarus; CNPq and FAPESP,
Brazil; NSERC, NRC and CFI, Canada; CERN; CONICYT, Chile; CAS, MOST and
NSFC, China; COLCIENCIAS, Colombia; MSMT CR, MPO CR and VSC CR, Czech
Republic; DNRF, DNSRC and Lundbeck Foundation, Denmark; ARTEMIS, European
Union; IN2P3-CNRS, CEA-DSM/IRFU, France; GNAS, Georgia; BMBF, DFG, HGF, MPG
and AvH Foundation, Germany; GSRT, Greece; ISF, MINERVA, GIF, DIP and
Benoziyo Center, Israel; INFN, Italy; MEXT and JSPS, Japan; CNRST, Morocco;
FOM and NWO, Netherlands; RCN, Norway; MNiSW, Poland; GRICES and FCT,
Portugal; MERYS (MECTS), Romania; MES of Russia and ROSATOM, Russian
Federation; JINR; MSTD, Serbia; MSSR, Slovakia; ARRS and MVZT, Slovenia;
DST/NRF, South Africa; MICINN, Spain; SRC and Wallenberg Foundation,
Sweden; SER, SNSF and Cantons of Bern and Geneva, Switzerland; NSC, Taiwan;
TAEK, Turkey; STFC, the Royal Society and Leverhulme Trust, United Kingdom;
DOE and NSF, United States of America.

The crucial computing support from all WLCG partners is acknowledged
gratefully, in particular from CERN and the ATLAS Tier-1 facilities at
TRIUMF (Canada), NDGF (Denmark, Norway, Sweden), CC-IN2P3 (France),
KIT/GridKA (Germany), INFN-CNAF (Italy), NL-T1 (Netherlands), PIC (Spain),
ASGC (Taiwan), RAL (UK) and BNL (USA) and in the Tier-2 facilities
worldwide.

\,
\,

\clearpage
\begin{flushleft}
{\Large The ATLAS Collaboration}

\bigskip

G.~Aad$^{\rm 48}$,
B.~Abbott$^{\rm 111}$,
J.~Abdallah$^{\rm 11}$,
A.A.~Abdelalim$^{\rm 49}$,
A.~Abdesselam$^{\rm 118}$,
O.~Abdinov$^{\rm 10}$,
B.~Abi$^{\rm 112}$,
M.~Abolins$^{\rm 88}$,
H.~Abramowicz$^{\rm 153}$,
H.~Abreu$^{\rm 115}$
% ...
\bigskip

% ...
$^{10}$ Institute of Physics, Azerbaijan Academy of Sciences, Baku, Azerbaijan\\
$^{11}$ Institut de F\'isica d'Altes Energies and Departament de F\'isica de la Universitat Aut\`onoma  de Barcelona and ICREA, Barcelona, Spain\\
% ...
$^{48}$ Fakult\"{a}t f\"{u}r Mathematik und Physik, Albert-Ludwigs-Universit\"{a}t, Freiburg i.Br., Germany\\
$^{49}$ Section de Physique, Universit\'e de Gen\`eve, Geneva, Switzerland\\
% ...
$^{88}$ Department of Physics and Astronomy, Michigan State University, East Lansing MI, United States of America\\
% ...
$^{111}$ Homer L. Dodge Department of Physics and Astronomy, University of Oklahoma, Norman OK, United States of America\\
$^{112}$ Department of Physics, Oklahoma State University, Stillwater OK, United States of America\\
% ...
$^{115}$ LAL, Univ. Paris-Sud and CNRS/IN2P3, Orsay, France\\
% ...
$^{118}$ Department of Physics, Oxford University, Oxford, United Kingdom\\
% ...
$^{153}$ Raymond and Beverly Sackler School of Physics and Astronomy, Tel Aviv University, Tel Aviv, Israel\\
% ...
\end{flushleft}


\begin{thebibliography}{9}
\bibitem{ATLASCoordinateSystem} The ATLAS reference system is a
Cartesian right-handed coordinate system, with the nominal collision
point at the origin.  The counter-clockwise beam direction defines the
positive $z$-direction, while the positive $x$-direction is defined as
pointing from the collision point to the center of the LHC ring and
the positive $y$-axis points upwards. The azimuthal angle $\phi$ is
measured around the beam axis and the polar angle $\theta$ is measured
with respect to the $z$-axis. The pseudorapidity is defined as $\eta =
- \ln \left[ \tan(\theta/2) \right]$, while the rapidity is defined as
$y=\frac{1}{2} \ln \frac{E+p_L}{E-p_L}$, where $E$ is the particle
energy and $p_L$ is the particle momentum along the $z$-axis.
\end{thebibliography}
\end{document}